\documentclass[10pt,aps,twocolumn,showkeys,showpacs,preprintnumbers,prd,superscriptaddress,nofootinbib]{revtex4-2}
\usepackage{amsmath}

\usepackage{amsfonts}
\usepackage{amssymb}

\usepackage{natbib}
\usepackage{bm}     
\usepackage{booktabs}
\usepackage{dcolumn}   
\usepackage{graphicx}
\usepackage{epstopdf}
\usepackage{mdwmath}   
\usepackage{booktabs}  
\usepackage{lipsum}
\usepackage{multirow}
\usepackage{soul}     
\usepackage{verbatim}  
\usepackage{eqparbox}  
\usepackage{xcolor}
\usepackage{url}       
\usepackage{csquotes}  
\usepackage{blindtext} 
\usepackage{rotating}  
\usepackage{adjustbox} 
\usepackage{hyperref}  
\usepackage{tabularx}
\usepackage{array}
\usepackage{orcidlink}

\usepackage[capitalize]{cleveref}

\definecolor{amethyst}{rgb}{0.6, 0.4, 0.8}
\definecolor{customblue}{rgb}{0.2, 0.2, 0.8}  
\hypersetup{
    linktoc=all,
    colorlinks=true,
    linkcolor=blue,
    citecolor=red,
    urlcolor=magenta
}

\makeatletter\let\expandableinput\@@input\makeatother

\graphicspath{{images/}}

\newcommand{\be}{\begin{equation}}
\newcommand{\ee}{\end{equation}}
\newcommand{\ba}{\begin{eqnarray}}
\newcommand{\ea}{\end{eqnarray}}

\begin{document}

\title{Dissipative Cosmology and the Nature of Dark Energy: Insights from Bulk Viscosity with DESI DR2 observations}

\author{Shahnawaz A. Adil\,\orcidlink{0000-0003-4999-7801}}
\email{shahnawaz@icf.unam.mx}
\affiliation{Instituto de Ciencias F\'isicas, Universidad Nacional Aut\'onoma de M\'exico, Cuernavaca, 
Morelos, 62210, M\'exico}
\author{Sonej Alam\,\orcidlink{0009-0008-8322-2923},}
\email{sonejalam36@gmail.com}
\affiliation{Department of Physics, Jamia Millia Islamia, New Delhi 110025, India}
\author{Somasri Sen\,\orcidlink{0000-0001-8449-7242},}
\email{ssen@jmi.ac.in}
\affiliation{Department of Physics, Jamia Millia Islamia, New Delhi 110025, India}
\author{J. Alberto Vazquez\,\orcidlink{0000-0002-7401-0864}}
\email{javazquez@icf.unam.mx}
\affiliation{Instituto de Ciencias F\'isicas, Universidad Nacional Aut\'onoma de M\'exico, Cuernavaca, Morelos, 62210, M\'exico}

\begin{abstract}
We explore a cosmological model in which dark energy is described by a bulk viscous fluid, providing a dissipative mechanism for late-time cosmic acceleration. Considering both minimally and non-minimally coupled scenarios, we constrain the model using SNe Ia, DESI DR2 BAO, and Planck 2018 CMB data. We find that viscous effects can successfully mimic dynamical dark energy and yield improved fits over $\Lambda$CDM, particularly in the interacting non-minimal case. Our results demonstrate that dissipative processes offer a viable and physically motivated alternative to the cosmological constant in explaining the current accelerated expansion of the universe.
\end{abstract}

\maketitle
\section{Introduction}

The last decade of the twentieth century witnessed a profound transformation in our understanding of the universe. The pioneering observations from Type Ia supernovae \cite{1} in the late 1990s revealed that the universe is, in fact, expanding at a faster rate. This discovery, supported by data from cosmic microwave background (CMB) \cite{2} measurements to baryon acoustic oscillations (BAO) \cite{eBOSS:2020yzd}, marked a paradigm shift in cosmology. It became evident that the dominant component of the universe's energy budget is not dark matter but a mysterious entity termed dark energy \cite{2,4,5,6,7,1,eBOSS:2020yzd}, which exerts a negative pressure capable of overcoming the gravitational pull of dark matter and driving the accelerated expansion. Today, the dark sector—comprising dark matter (approximately 26\%) and dark energy (approximately 69\%) accounts for about 95\% of the total energy budget of the universe, with the remaining 5\% consisting of ordinary baryonic matter \cite{2}. 
 
The simplest and most widely accepted framework to model this dark universe is the $\Lambda$CDM model \cite{2}, where $\Lambda$ is the cosmological constant representing dark energy and CDM represents dark matter. This model has been remarkably successful in explaining a wide range of cosmological observations \cite{2}, from the large-scale structure of the universe to the precise measurements of the CMB. However, despite its successes, the $\Lambda$CDM model has both theoretical and observational concerns. Two of the most significant theoretical issues are the coincidence problem \cite{2014EPJC...74.3160V,2021arXiv210708916Z} and the fine-tuning problem \cite{2022RSPTA.38010182S,2007JPhA...40.6583S}. On the other hand, high-precision recent cosmological observations have revealed significant cosmological tensions that challenge the $\Lambda$CDM model. The present value of the Hubble constant $H_0$, measured from the local universe using Type Ia supernovae and Cepheid variables \cite{2020ApJ...889....5H,2021ApJ...911...65B,2023ApJ...953...35G,2022MNRAS.514.4620D,2020ApJ...891L...1P,2020ApJ...896....3K,2018MNRAS.474.1250F,2024SSRv..220...48B,2022LRR....25....6M,2023ApJ...956..111W,2023ApJ...944...94T,2023MNRAS.523.2369S,2021ApJ...919...16F,2022ApJ...932...15A,2023A&A...673A...9S,2023ApJ...954L..31S,2024ApJ...963L..43A}, differs from the value inferred from early-universe CMB-data modeled with $\Lambda$CDM \cite{2022JHEAp..34...49A,2022NewAR..9501659P,2021APh...13102605D,2019NatAs...3..891V,2020PhRvD.101d3533K,2021CmPhy...4..123J,2021CQGra..38o3001D,2021A&ARv..29....9S,2023ARNPS..73..153K}, well known in the literature as the Hubble tension.  

Similarly, the mismatch between the amplitude of matter density fluctuations $S_8$ inferred from weak lensing surveys and galaxy clustering \cite{WL,WL2,WL3}, and that predicted by $\Lambda$CDM using CMB data \cite{2022JHEAp..34...49A,2021APh...13102605D}, is referred to as the $S_8$ tension. These tensions, along with the theoretical challenges of the $\Lambda$CDM model, have motivated physicists to explore alternative cosmological models that may resolve these discrepancies and provide a more natural explanation for the observed phenomena.

Alternative models can be broadly classified into two categories based on their approach to modifying Einstein's field equations. The first category involves modifications to the geometry, such as DGP braneworld models \cite{2000PhLB..485..208D,2001PhLB..502..199D,2002PhRvD..65d4023D,2003JCAP...11..014S,2006PhRvD..73d3518Z}, holographic dark energy models \cite{2004PhLB..603....1L,2006GReGr..38.1285N,2009PhRvD..79d3511G}, and $f(R)$ gravity \cite{2004PhRvD..70d3528C,2007PhRvD..76f4004H,2006PhRvD..74h6005N,2010LRR....13....3D, Alam:2024kgk}. These models attempt to explain recent cosmological dynamics by altering the gravitational sector itself. The second category focuses on modifications to the matter part, where different kinds of components suitable for explaining the dynamics are explored. This approach can be further divided into three subcategories: non-interacting dark energy models, interacting dark energy models, and unified dark matter (UDM) models. Non-interacting models \cite{Ratra:1987rm,Peebles:2002gy,2004PhLB..603....1L,2001PhLB..511..265K} treat dark matter and dark energy as distinct entities that evolve independently, while interacting models \cite{2000PhRvD..62d3511A,2005PhLB..624..141W} allow for energy exchange between the two components. UDM models, on the other hand, propose that dark matter and dark energy are different manifestations of a single fluid, sometimes characterized by an equation of state of the form $p=f(\rho)$. This approach has been successfully applied in models like the Chaplygin gas \cite{2001PhLB..511..265K,2002astro.ph..7430F}, generalized Chaplygin gas \cite{2002dmap.conf..306B,2002PhRvD..66d3507B}, and modified Chaplygin gas \cite{2002hep.th....5140B,2011BrJPh..41..333D}. 

More recently, observations from DESI have further intensified interest in dynamical dark-energy scenarios~\cite{DESI:2024mwx}. In particular, several analyses based on DESI BAO measurements have indicated mild preferences for deviations from a pure cosmological constant, especially when combined with other low-redshift probes \cite{Gomez-Valent:2021cbe,Gomez-Valent:2022bku,Giare:2024smz,Berghaus:2024kra,Wolf:2025jlc,Lee:2025pzo,Zhong:2025gyn,Yao:2025wlx,Qu:2024lpx,Wang:2024dka,Giare:2024gpk,Gialamas:2024lyw,Shlivko:2024llw,Ye:2024ywg,Akthar:2024tua,Chan-GyungPark:2025cri,Ferrari:2025egk,Sohail:2024oki,Alam:2025epg,Sohail:2025mma,Hossain:2025gpr}. These developments have renewed attention toward late-time dynamical dark-energy models capable of modifying the post-recombination expansion history without altering the physics of the early Universe.

A common phenomenological approach to introducing dynamics in the DE sector is to promote the cosmological-constant EoS to a redshift-dependent quantity, $\omega_{\rm de}=\omega_{\rm de}(z)$. This allows departures from the standard $\Lambda$CDM evolution while maintaining a relatively simple effective-fluid description. A wide variety of parameterizations have been proposed in the literature, including the linear parametrization~\cite{Akarsu:2015yea, cooray1999gravitational,ASTIER20018,PhysRevD.65.103512}, logarithmic parametrization~\cite{Efstathiou:1999tm}, the Chevallier-Polarski-Linder (CPL) parametrization~\cite{chevallier:hal-00142125, CPL}, the Jassal-Bagla-Padmanabhan (JBP)~\cite{jassal}, and the Barboza-Alcaniz (BA)~\cite{Barboza_2008} parametrizations. Beyond these phenomenological constructions, more fundamental theoretical frameworks have also been explored, including Quintessence~\cite{quin1,quin2,quin3}, Phantom~\cite{phant1,phant2, Vazquez:2020ani}, K-essence~\cite{kessense1,kessense2,kessense3}, Quintom models~\cite{Vazquez:2023kyx, quintom1,quintom2,quintom3}, interacting dark-energy scenarios~\cite{Garcia-Arroyo:2024tqq,DiValentino:2017iww,DiValentino:2019ffd,Pan:2019jqh,DiValentino:2020kha,Giare:2024bxz,Anchordoqui:2021ghd}, and early dark-energy models~\cite{early1,early2,early3,early4}. 

Despite the extensive research conducted in these directions, the origin, nature, and evolution of the dark components remain elusive. This has led to a growing interest in dissipative processes, particularly viscosity, as a potential mechanism to explain cosmological phenomena \cite{2017EPJC...77..660W,PhysRevD.101.044010}. Dissipative processes play a crucial role in the thermodynamic evolution of the universe. Viscosity, which arises from deviations from local thermodynamic equilibrium, can be of two types: shear viscosity and bulk viscosity. In cosmology, shear viscosity is often neglected due to the large-scale isotropy of the universe. Bulk viscosity, on the other hand, can emerge from rapid changes in the expansion rate of the cosmic fluid, creating an effective pressure that drives the system back to equilibrium \cite{2020EPJP..135..718H,2016arXiv161001827B}. Once equilibrium is restored, the viscous pressure dissipates. This mechanism allows bulk viscosity to act as a natural source of acceleration without invoking exotic forms of matter. Hence, for obvious reasons, bulk viscous phenomena are relevant in both the early and late stages of the evolution of the universe \cite{2017IJMPD..2630024B,2005MPLA...20.1729C,2023EPJC...83.1155Y,2020Symm...12.1085B}. In the context of early-universe inflation, bulk viscosity has been proposed as a mechanism to drive the rapid expansion required to solve the horizon and flatness problems \cite{2016arXiv160403873V,2012JCAP...11..042B,1995GReGr..27..171V}. The dissipative nature of viscous fluids can naturally lead to entropy production and particle creation, providing a plausible explanation for the thermal history of the universe \cite{2000PhRvD..61h3511Z,Triginer:1994nq}. Similarly, in the late universe, bulk viscosity has been explored as a potential driver of the observed accelerated expansion, offering an alternative to the cosmological constant and other dark energy models \cite{2010JCAP...08..009A,PhysRevLett.114.091301}. Bulk viscosity introduces dissipative effects into the cosmic fluid, which can modify the dynamics of the universe by generating an effective negative pressure \cite{2022Univ....9...12S}. This effective pressure can counteract gravitational collapse and drive acceleration without invoking exotic scalar fields or modifications to general relativity. Viscous models are particularly appealing as they naturally arise in the context of imperfect fluids, providing a physically motivated mechanism to explain late-time cosmic acceleration \cite{2022IJGMM..1950060T,2021NewA...8201452A,2017IJMPD..2630024B}. Viscous fluid has also been explored as a single fluid unifying  both the dark sector— dark matter and dark energy.
In such unified dark matter (UDM) models, the dark components are not treated as separate entities but as different manifestations of a single viscous fluid \cite{2010JCAP...01..014P,PhysRevD.110.063513}. 

In recent years, viscous models have gained renewed attention due to their ability to address some of the challenging issues in cosmology. Viscous effects may act as an effective dark energy component, mimicking the late-time accelerated expansion without requiring a cosmological constant \cite{2020arXiv200614153T,2023EPJC...83.1155Y}. These features make viscous models an attractive framework for exploring deviations from the standard $\Lambda$CDM paradigm while remaining consistent with observational constraints. Viscous models can alter the expansion history of the universe, potentially reconciling the discrepancy between different observational datasets. For instance, bulk viscosity has been proposed as a mechanism to explain the observed tension in the Hubble constant ($H_{0}$) between local measurements (e.g., SH0ES) and those inferred from the CMB (e.g., Planck) \cite{2021MPLA...3650198N,2017EPJC...77..660W,2025PDU....4801847M}.  Studies are carried out on growth of cosmic structures, providing insights into the formation of galaxies and large-scale structures in the presence of dissipative processes \cite{2018arXiv180506363A,2024arXiv241202276M,2017PhRvD..96b3527B}. They also provide an avenue to address the $S_8$ tension by modifying the rate of structure growth through viscosity-induced damping of matter density fluctuations~\cite{2017JCAP...11..005A,2023ApJ...959..120A,2018MNRAS.481.1799M}.

In this work, we investigate the role of bulk viscosity in the evolution of the universe by considering two distinct scenarios: minimal coupling and non-minimal coupling between the dark components. In the minimal coupling scenario, the dark components evolve independently, following the standard conservation laws. In contrast, the non-minimal coupling scenario introduces an interaction between dark matter and dark energy, parameterized by a coupling constant. The background framework for this model is outlined in Section \ref{sec:model}. The data analysis is described in Section \ref{sec:data}, while the results are presented and discussed in Section~\ref{sec:results}. Finally, the conclusions and implications of this study are summarized in Section~\ref{sec:conclusion}.

\section{Cosmological Field Equations for the Dual Fluid Model} \label{sec:model}

In this work, we consider the Universe to be composed of three main components: radiation, matter, and dark energy. Radiation dominates the early Universe, while matter and dark energy become important during the later stages of cosmic evolution. Each component is modeled as a distinct fluid: radiation is treated as a relativistic perfect fluid, matter as a non-relativistic fluid, and dark energy as a viscous fluid containing both thermodynamic and dissipative pressure contributions. The total energy-momentum tensor is given by
\begin{equation}
T_{\mu\nu} = \sum_i (\rho_i + p_i) u_\mu u_\nu + p_i g_{\mu\nu},
\end{equation}
where $i=r,m,d$ correspond to radiation, matter, and dark energy, respectively. Here, $\rho_r$, $\rho_m$, and $\rho_d$ denote the energy densities of radiation, matter, and dark energy, while $p_r$, $p_m$, and $p_d$ are their corresponding pressures. In a comoving frame, the four-velocity satisfies $u^\mu=\delta^\mu_0$ and $u_\mu u^\mu=-1$.

The matter sector is assumed to consist of baryonic matter and dark matter, such that $\rho_m=\rho_b+\rho_{dm}$. We assume that baryons are pressureless, $p_b=0$, whereas the dark matter component obeys the barotropic relation $p_{dm}=\gamma \rho_{dm}$, where $\gamma$ is the dark matter equation-of-state parameter. Radiation satisfies the standard relation $p_r=\rho_r/3$.

The pressure associated with the viscous dark-energy fluid consists of two contributions: a thermodynamic pressure and a bulk viscous pressure, such that $p_d=p_t+\pi$. The thermodynamic part is parameterized as $p_t=\omega \rho_d$, where $\omega$ is the equation-of-state parameter of the viscous fluid. The viscous contribution is given by $\pi= \eta u^\mu_{;\mu}$, where $\eta$ is the bulk viscosity coefficient and $u^\mu_{;\mu}$ is the expansion scalar. Consequently, the effective pressure of the viscous fluid becomes
\begin{equation}
\label{pd}
p_d = \omega \rho_d - \eta u^\mu_{;\mu}.
\end{equation}

We note that the bulk viscosity coefficient is assumed to be positive definite, $\eta>0$. In general, $\eta$ may either be treated as a constant or parameterized as a function of the energy density, i.e. $\eta\equiv\eta(\rho)$.

We assume a flat, isotropic, and homogeneous background, described by the Friedmann-Robertson-Walker (FRW) metric:
\begin{equation}
ds^2 = -dt^2 + a^2(t) \bigg[dr^2 + r^2 \big(d\theta^2 + \sin^2\theta \, d\phi^2\big)\bigg],
\end{equation}
where $a(t)$ is the scale factor of the universe. With this background, the dynamical field equations that govern the cosmological evolution are:
\begin{eqnarray}
3H^2 &=& \sum_i \rho_i, \label{hsq} \\
2\dot{H} + 3H^2 &=& - \sum_i p_i, \label{hdot}
\end{eqnarray}
where $H = \frac{\dot{a}}{a}$ is the Hubble parameter. Here, an overhead dot represents a derivative with respect to time. For simplicity, we have adopted the natural units $8\pi G = c = 1$. To explore the role of viscosity in the evolution of the universe, we consider two cases: minimal interaction and non-minimal interaction.

\subsection{Minimal Interaction}

In the minimal interaction scenario, the different components of the Universe (radiation, matter, and viscous dark energy), evolve independently and satisfy the standard conservation laws. Consequently, the energy-momentum tensor of each component is separately conserved, leading to the continuity equation
\begin{equation}
\dot{\rho}_i + 3H(\rho_i+p_i)=0,
\label{droi}
\end{equation}
where $i=r,m,d$ corresponds to radiation, matter, and viscous dark energy, respectively.

Using the corresponding equations of state in equation~\eqref{droi}, the radiation energy density evolves as $\rho_r=\rho_{r0}a^{-4}$, where $\rho_{r0}$ is the present-day radiation energy density. The matter sector is composed of baryonic matter and dark matter, such that $\rho_m=\rho_b+\rho_{dm}$. We assume baryons to be pressureless, $p_b=0$, while the dark matter component satisfies the barotropic relation $p_{dm}=\gamma \rho_{dm}$, where $\gamma$ denotes the dark matter equation-of-state parameter. Under these assumptions, the baryon and dark matter densities evolve as $\rho_b=\rho_{b0}a^{-3}$ and $\rho_{dm}=\rho_{dm0}a^{-3(1+\gamma)}$, respectively. For the continuity equation of the viscous fluid, the effective pressure ($p_d$) is given in equation (\ref{pd}). We assume the bulk viscosity coefficient $\eta$ varies as a simple power law of the dark energy density, i.e., $\eta = \eta_0 \rho_d^n$, where $\eta_0$ and $n$ are free parameters. This form of $\eta$ has been widely adopted in the literature \cite{2023IJGMM..2050086S,Singh:2008zzj,2012mgm..conf.1353C}. Various other forms of $\eta$ are also available in the literature \cite{2009JCAP...04..006A, 2010A}. Thus, the conservation equation for the viscous fluid can be rewritten as:
\begin{equation}
\dot{\rho_d} + 3H\big[(1+w)\rho_d - 3\eta_0 H \rho_d^n\big] = 0.
\end{equation}
Combining the conservation equations  with the field equations, one can arrive at the following evolution equation for the Hubble parameter:
\begin{eqnarray}
\label{gent}
2\dot{H} + 3H^2(1+w) + (\gamma - \omega)\rho_m - \bigg(\omega - \frac{1}{3}\bigg)\rho_r, \nonumber\\
- 3^{n+1}\eta_0 H^{2n+1} + n 3^n \eta_0 H^{2n-1}(\rho_m + \rho_r) = 0.
\end{eqnarray}

The model has four key parameters: $\gamma$, $\omega$, $n$, and $\eta_0$. Among these, $\gamma$, $\omega$, and $n$ are dimensionless constants and are not subject to specific physical constraints. $\eta_0$ is positive. If $n=\frac{1}{2}$, $\eta_0$ is a dimensionless positive definite constant. But if $n \neq \frac{1}{2}$, then $\eta_0$  is a dimensionful positive definite constant. To better understand the behavior of the model, we first discuss the case of $n=\frac{1}{2}$, followed by the more general case where $n$ can take any value.

For $n = \frac{1}{2}$, the above equation (\ref{gent}) simplifies to:
\begin{eqnarray}
2\dot{H} + 3\bigg[(1+w) - \sqrt{3}\eta_0\bigg] H^2 - \bigg[\omega - \frac{1}{3}- \frac{\sqrt{3}}{2}\eta_0\bigg]\rho_r \nonumber\\
+ \bigg[\gamma - \omega + \frac{\sqrt{3}}{2}\eta_0\bigg]\rho_m = 0.
\end{eqnarray}

It is straightforward to derive an analytical solution to this first order ordinary differential equation:
\begin{align}
\label{halfh2}
\frac{H^2}{H_0^2} =  
&\frac{\omega-\frac13-\frac{\sqrt3\eta_0}{2}}
{\omega-\frac13-\sqrt3\eta_0}
\,\Omega_{r0}a^{-4}
+
\frac{\gamma-\omega+\frac{\sqrt3\eta_0}{2}}
{\gamma-\omega+\sqrt3\eta_0}
\,\Omega_{m0}a^{-3(1+\gamma)}
\nonumber\\
&+
\Bigg(
1
-
\frac{\omega-\frac13-\frac{\sqrt3\eta_0}{2}}
{\omega-\frac13-\sqrt3\eta_0}\Omega_{r0}
-
\frac{\gamma-\omega+\frac{\sqrt3\eta_0}{2}}
{\gamma-\omega+\sqrt3\eta_0}\Omega_{m0}
\Bigg)
\nonumber\\
&\times a^{-3(1+\omega-\sqrt3\eta_0)}
.
\end{align}
where $H_0$ is the present Hubble parameter, and $\Omega_{m0} = \frac{\rho_{m0}}{3H_0^2}$ and $\Omega_{r0} = \frac{\rho_{r0}}{3H_0^2}$ are the dimensionless matter and radiation density parameters, respectively. 

Interestingly, equation (\ref{halfh2}) is mathematically identical to the case of a dark energy model with a constant equation of state, provided we redefine:
\[
\tilde{\Omega}_{r0} = \frac{\omega - \frac{1}{3} - \frac{\sqrt{3}\eta_0}{2}}{\omega - \frac{1}{3} - \sqrt{3}\eta_0}\,\Omega_{r0}, \quad \tilde{\Omega}_{m0} = \frac{\gamma - \omega + \frac{\sqrt{3}\eta_0}{2}}{\gamma - \omega + \sqrt{3}\eta_0}\,\Omega_{m0}.
\]
Since most observational analyses depend primarily on the background cosmology, all analyses for the constant equation of state scenario are directly applicable here.

Now, let us consider the more general case where $n$ can take any value. In this case, to maintain dimensional consistency for the viscous pressure, $\eta_0$ becomes a dimensionful quantity. For the sake of statistical analysis, we redefine $\eta_0$ in terms of  a dimensionless parameter $\alpha$ by defining $\alpha = \eta_0 \rho_{0}^{n - \frac{1}{2}}$, $\rho_0$ is the total energy density at present. Thus, the model parameters are now $\omega$, $\gamma$, $n$, and $\alpha$. Rewriting equation (\ref{gent}) in terms of the new dimensionless parameter $\alpha$ and the normalized Hubble parameter we obtain:
\begin{align}
\label{eq:diff_hubble_minimal}
a\frac{dE}{da}
&+ \frac{3}{2}(1+w)E
+ \frac{3}{2}(\gamma-\omega)
\frac{\Omega_{m0}a^{-3(1+\gamma)}}{E}
\nonumber\\
&- \frac{3}{2}\left(\omega-\frac{1}{3}\right)
\frac{\Omega_{r0}a^{-4}}{E}
\nonumber\\
&- \frac{3\sqrt{3}}{2}\alpha E^{2n}
\Bigg[
1-
\frac{
\Omega_{m0}a^{-3(1+\gamma)}
+\Omega_{r0}a^{-4}
}{E^2}
\Bigg]^n
=0 .
\end{align}
where ${E} = H/H_0$ is the normalized Hubble parameter.

We assume that the equations of state for baryons and dark matter are zero and $\gamma$, respectively. Under this assumption, equation \eqref{eq:diff_hubble_minimal} reduces to

\begin{align}
\label{eq:hubble2}
a\frac{dE}{da}
&+ \frac{3}{2}(1+w)E
+ \frac{3}{2}(\gamma-\omega)
\frac{\Omega_{dm0}a^{-3(1+\gamma)}}{E}
- \frac{3}{2}\omega
\frac{\Omega_{b0}a^{-3}}{E}
\nonumber\\
&- \frac{3}{2}\left(\omega-\frac13\right)
\frac{\Omega_{r0}a^{-4}}{E}-\frac{3\sqrt3}{2}\alpha E^{2n}
\nonumber\\
&\Bigg[
1-
\frac{
\Omega_{dm0}a^{-3(1+\gamma)}
+\Omega_{b0}a^{-3}
+\Omega_{r0}a^{-4}
}{E^2}
\Bigg]^n
=0 .
\end{align}

This first-order nonlinear differential equation cannot be solved analytically. We solve it numerically which is elaborated in detail in appendix~\ref{appendix:a}. 

\subsection{Non-Minimal Interaction}

In this case, we consider an interaction between matter and the viscous fluid, implying that the total energy-momentum tensor of these two components obeys a combined conservation relation. The interaction parameter $\xi$ introduces a coupling between matter and the viscous fluid, regulating the transfer of energy between the two sectors. Such interactions have been widely explored as a possible extension of the standard cosmological model, since they can modify the late-time evolution of the Universe, generate departures from the standard $\Lambda$CDM scenario, and potentially alleviate ongoing cosmological tensions, including those associated with the measurements of $H_0$ and $S_8$. Considering the interaction, the conservation equations for these fluids obey: 

\begin{eqnarray}
\dot{\rho}_{dm} + 3H(\rho_m+p_m) = \xi H \rho_{dm},\label{drom1}\\
\dot{\rho_d} + 3H(\rho_d+p_d) = -\xi H \rho_{dm}.\label{drod1}
\end{eqnarray}
\noindent
Here, $\xi$ is the interaction parameter that governs the energy transfer between the matter and viscous fluid. A positive value of $\xi$ indicates energy transfer from matter to the viscous fluid, while a negative value indicates the opposite. 

Equation (\ref{drom1}) implies that the energy density of matter scales as:
\begin{equation}
    \rho_{dm} = \rho_{dm0} a^{-3(1+\gamma)+\xi}.
\end{equation}
In terms of the effective pressure of the viscous fluid,
\begin{equation}
p_d = w\rho_d - 3\eta_0 H \rho_d^n, \label{eq:pressure_de}
\end{equation}
the energy density of the dissipative fluid evolves as follows:
\begin{equation}
\dot{\rho}_d + 3H\left[(1+w)\rho_d - 3\eta_0 H \rho_d^n \right] = -\xi H \rho_{dm}.
\end{equation}
Rewriting the above equation in terms of the dimensionless constant $\alpha$, and finally, by combining it with others, the evolution equation of the dimensionless normalised Hubble parameter can be written as:
\begin{align}
\label{eq:hubble_non-minimal}
a\frac{dE}{da}
&+ \frac{3}{2}(1+w)E
+ \frac{3}{2}(\gamma-\omega)
\frac{\Omega_{dm0}a^{-3(1+\gamma)+\xi}}{E}
- \frac{3}{2}\omega
\frac{\Omega_{b0}a^{-3}}{E}
\nonumber\\
&- \frac{3}{2}\left(\omega-\frac13\right)
\frac{\Omega_{r0}a^{-4}}{E}- \frac{3\sqrt3}{2}\alpha E^{2n}
\nonumber\\
&\Bigg[
1-
\frac{
\Omega_{dm0}a^{-3(1+\gamma)+\xi}
+\Omega_{b0}a^{-3}
+\Omega_{r0}a^{-4}
}{E^2}
\Bigg]^n
=0 .
\end{align}
At this stage, the model parameters are $\gamma$, $\omega$, $n$, $\alpha$, and $\xi$. These parameters will be constrained using observational data in subsequent sections.  We explore the implications of varying $\xi$ on the model's dynamics and its consistency with observations.

\section{Data and Analysis} \label{sec:data}

To analyze the model and obtain constraints on its parameters, we have combined different recent observational data.  We have taken into account baryon acoustic oscillations (BAO), cosmic microwave background (CMB) and SNe Ia samples from Panthen+\&SH0ES and Union3. 
\begin{enumerate}

\item \textbf{Union3:} The updated `Union' compilation of 2,087 cosmologically useful SNe Ia from 24 datasets ('Union3'). These 2,087 SNe Ia are then compressed to 22 redshift bins. In our analysis, we use these binned modulus distance observations of Union3 compilations \cite{Rubin:2023jdq}. 

\item \textbf{Pantheon+\&SH0ES :} We include 1701 distance modulus measurements of Type Ia supernovae from the Pantheon+ sample~\cite{Brout:2022vxf}, consisting of 1,550 light curves over the redshift range \texttt{$0.001 \leq z \leq 2.26$}. 

\item{\bf DESI BAO DR2}

We further incorporate BAO measurements from the Dark Energy Spectroscopic Instrument (DESI) data release 2 (DR2)~\cite{DESI:2025zgx}, which are derived from several classes of extragalactic tracers. These include: (i) the Bright Galaxy Sample (BGS), with over 5.5 million reliable redshifts in the range $0.1 < z < 0.4$; (ii) the Luminous Red Galaxy Sample (LRG), covering $0.4 < z < 0.6$ and $0.6 < z < 0.8$; (iii) the Emission Line Galaxy Sample (ELG), spanning $1.1 < z < 1.6$; (iv) the combined LRG and ELG sample (LRG+ELG), in the redshift range $0.8 < z < 1.1$; (v) the Quasar sample (QSO), extending from $0.8 < z < 2.14$; and (vi) the Lyman-$\alpha$ forest sample (Ly$\alpha$), probing the range $1.77 < z < 4.16$.

\item{\bf CMB}

The CMB constraint is incorporated using a compressed Gaussian likelihood based on the \textit{Planck 2018} data. Instead of the full likelihood, we use a reduced set of parameters that efficiently captures the relevant background information, namely the physical baryon density ($\omega_b$), baryon and cold dark matter density ($\omega_{bc}$), and the angular scale of the sound horizon ($\theta_*$). These quantities encode the key geometrical information through the ratio of the sound horizon to the angular diameter distance at recombination, thereby providing strong constraints on the expansion history. The corresponding theoretical predictions are computed from the cosmological model and compared to the observed values through a Gaussian likelihood. This compressed likelihood reproduces the main constraints of the full \textit{Planck} TT, TE, and EE data at the background level. We use the method mentioned in Appendix A of DESI DR2 \cite{DESI:2025zgx} for the likelihood implementation.

\end{enumerate}


\noindent
We perform the parameter estimation using the Bayesian inference framework, which is widely used in modern cosmological analyses. Within this approach, the posterior probability distribution of the model parameters is obtained from the product of the likelihood function and the prior probability distribution. Assuming Gaussian-distributed observational uncertainties, the likelihood function can be expressed as
\begin{equation}
{\cal L}(\Theta)\propto 
\exp\!\left[-\frac{\chi^2(\Theta)}{2}\right],
\end{equation}
where $\Theta$ denotes the set of free cosmological and model parameters. Throughout this work, we adopt uniform priors over the parameter space. Consequently, the posterior distribution becomes directly proportional to the exponential of the chi-square function, implying that smaller values of $\chi^2$ correspond to higher posterior probability and therefore better agreement with the observational data.

The total likelihood used in the analysis is constructed by combining the contributions from the different cosmological probes,
\begin{equation}
{\cal L}_{\rm tot}\propto 
\exp\!\left[-\frac{\chi^2_{\rm tot}}{2}\right],
\end{equation}
with
\begin{equation}
\chi^2_{\rm tot}=\chi^2_{\rm BAO}+\chi^2_{\rm CMB}+\chi^2_{\rm SN},
\end{equation}
where the supernova contribution corresponds either to the Union3 or the PantheonPlus+SH0ES compilation, depending on the dataset combination considered. This combined statistical analysis enables a consistent exploration of the parameter space and provides a robust assessment of the compatibility between different observational datasets within the viscous cosmological framework.

In our analysis, we constrain several cosmological and phenomenological parameters. The cosmological sector includes the present matter density parameter $\Omega_m$, the physical baryon density $\Omega_b h^2$, and the dimensionless Hubble parameter
\begin{equation}
h \equiv \frac{H_0}{100\,{\rm km\,s^{-1}\,Mpc^{-1}}}.
\end{equation}
The viscous dark-energy sector is characterized by the parameters $\alpha$, $w$, $\gamma$, and $n$, corresponding respectively to the dimensionless viscosity coefficient, the dark-energy equation-of-state parameter, the dark matter equation-of-state parameter, and the scaling index of the viscosity term. In the non-minimal interaction scenario, an additional coupling parameter $\xi$ is introduced to describe the interaction between dark matter and the viscous dark-energy sector.

The Bayesian parameter estimation and evidence calculations are performed using the dynamic nested-sampling algorithm implemented in the \texttt{dynesty} package~\cite{Speagle_2020}, integrated within the \texttt{SimpleMC} framework~\cite{BOSS:2014hhw,Padilla:2019mgi}. The resulting one- and two-dimensional posterior distributions are analyzed and visualized using the \texttt{GetDist} package~\cite{Lewis:2019xzd}. In addition, the functional reconstructions of $\rho(z)$, $\rho(z)+p(z)$, and the DESI DR2 observables presented in the following section are obtained using the \texttt{fgivenx} package~\cite{handley2019fgivenx}. The flat priors adopted for all model parameters are summarized in Table~\ref{tab:priors_visc}.

\begin{table}[htbp]
\centering
\begin{tabular}{lc}
\hline\hline
Parameter  & Prior range \\
\hline
$\Omega_m$          & $[0.2,\;0.8]$ \\
$\Omega_b h^2$     & $[0.021,\;0.024]$ \\
$h$                 & $[0.6,\;0.8]$ \\
$w$                 & $[-2.5,\;-0.33]$ \\
$\gamma$            & $[-0.3,\;0.3]$ \\
$\alpha$           & $[0.0,\;1.8]$ \\
$n$                & $[0.0,\;0.8]$ \\
$\xi$                & $[-0.3,\;1.5]$ \\
\hline\hline
\end{tabular}
\caption{Flat priors adopted for the cosmological and viscous model parameters in the analysis.}
\label{tab:priors_visc}
\end{table}


\section{Results and Discussion} \label{sec:results}

\begin{figure*}
    \centering
    \includegraphics[width=0.85\linewidth]{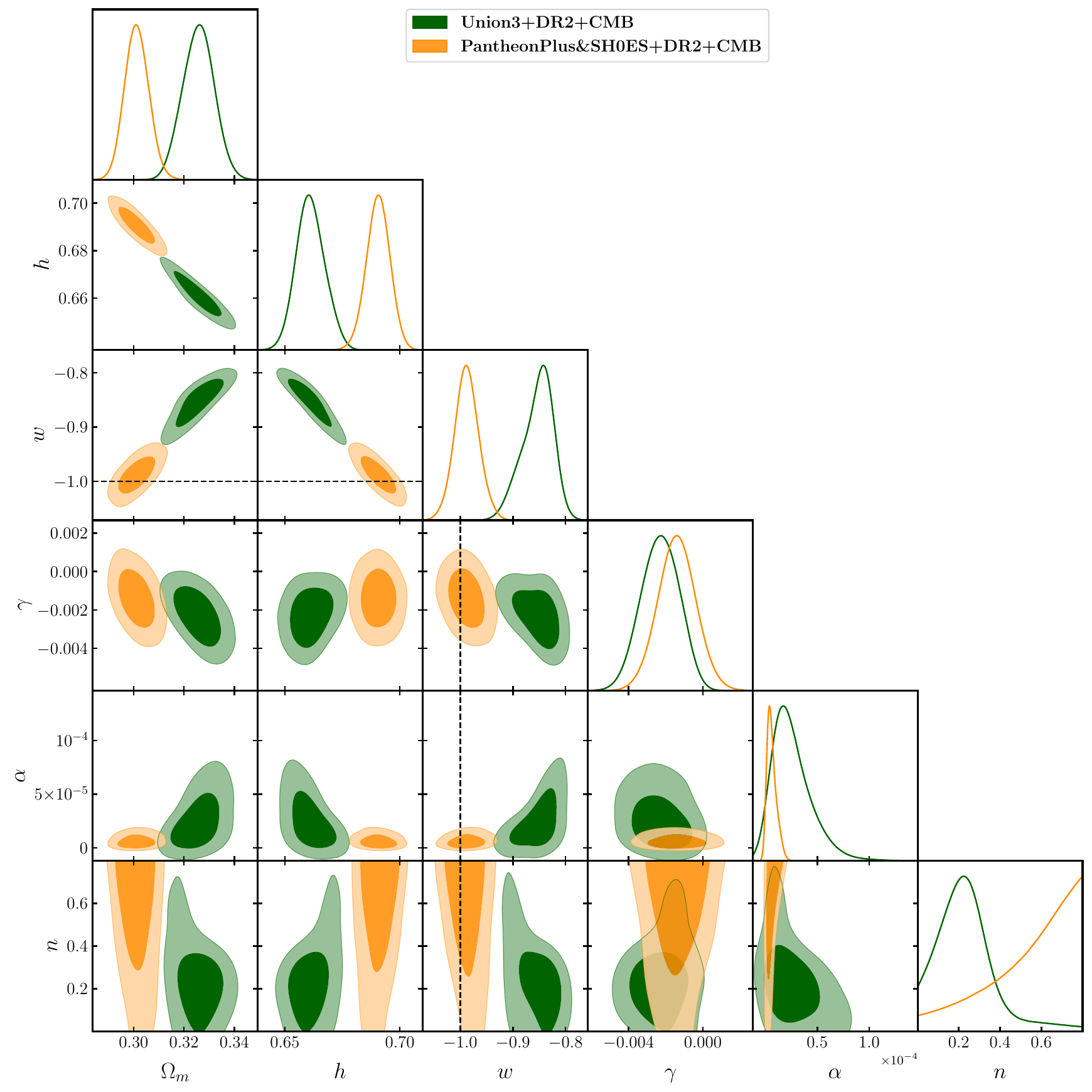}
    \caption{2-D posterior distributions with $2\sigma$ confidence levels for the minimal viscous scenario obtained using different SNe Ia datasets combined with DESI DR2 and Planck 2018 CMB constraints. The $\Omega_b h^2$ contours are omitted and presented in table~\ref{tab:combined_visc} since they are nearly identical in all cases owing to the strong BBN prior.}
    \label{fig:contour_minimal1}
\end{figure*}

\subsection{Case (i): Minimal Interaction}

\begin{table*}[htbp]
\centering
\renewcommand{\arraystretch}{1.3}
\resizebox{0.95\textwidth}{!}{
\begin{tabular}{l|cc|cc|cc}
\hline\hline

& \multicolumn{2}{c|}{$\Lambda$CDM} 
& \multicolumn{2}{c|}{Minimal} 
& \multicolumn{2}{c}{Non-minimal} \\

Parameter 
& U3 & P+ 
& U3 & P+ 
& U3 & P+ \\

\hline

$\Omega_m$        
& $0.3012 \pm 0.0037$ & $0.2971 \pm 0.0036$
& $0.3257 \pm 0.0062$ & $0.3012 \pm 0.0047$
& $0.3147 \pm 0.0066$ & $0.3012 \pm 0.0049$ \\

$\Omega_b h^2$    
& $0.02254 \pm 0.00011$ & $0.02263 \pm 0.00011$
& $0.02249 \pm 0.00014$ & $0.02237 \pm 0.00014$
& $0.02236 \pm 0.00014$ & $0.02236 \pm 0.00014$ \\

$h$               
& $0.6848 \pm 0.0031$ & $0.6885 \pm 0.0031$
& $0.6612^{+0.0056}_{-0.0067}$ & $0.6906 \pm 0.0051$
& $0.6747 \pm 0.0068$ & $0.6912 \pm 0.0052$ \\

$w$             
& --- & --- 
& $-0.853^{+0.035}_{-0.021}$ & $-0.988 \pm 0.023$
& $-0.515^{+0.16}_{-0.076}$ & $-0.57^{+0.19}_{-0.11}$ \\

$\gamma$          
& --- & --- 
& $-0.0023 \pm 0.0010$ & $-0.0014 \pm 0.0010$
& $0.065^{+0.025}_{-0.053}$ & $0.084^{+0.038}_{-0.063}$ \\

$\alpha$          
& --- & --- 
& $\left(2.5^{+1.0}_{-2.2}\right)\!\times\!10^{-5}$ 
& $\left(0.61^{+0.21}_{-0.57}\right)\!\times\!10^{-5}$
& $0.112^{+0.039}_{-0.029}$ & $0.116^{+0.050}_{-0.038}$ \\

$n$               
& --- & --- 
& $0.236^{+0.087}_{-0.13}$ & $0.543^{+0.25}_{-0.088}$
& $0.49^{+0.28}_{-0.14}$ & $0.46^{+0.28}_{-0.16}$ \\

$\xi$             
& --- & --- 
& --- & --- 
& $0.207^{+0.077}_{-0.17}$ & $0.27^{+0.12}_{-0.21}$ \\

\hline

$\mathcal{L}_{\mathrm{SN}}$
& $-15.24^{+0.20}_{-0.16}$ & $-790.8^{+1.3}_{-0.95}$
& $-13.87^{+0.82}_{-0.36}$ & $-789.0^{+1.6}_{-1.1}$
& $-13.14^{+0.52}_{-0.17}$ & $-788.1^{+1.6}_{-1.2}$ \\

$\mathcal{L}_{\mathrm{DR2}}$      
& $-5.98^{+0.90}_{-0.14}$ & $-5.51^{+0.42}_{-0.04}$
& $-11.0^{+2.7}_{-2.4}$ & $-6.46^{+1.3}_{-0.29}$
& $-5.12^{+0.97}_{-0.33}$ & $-6.41^{+1.6}_{-0.66}$ \\

$\mathcal{L}_{\mathrm{Planck18}}$ 
& $-2.40^{+1.3}_{-0.69}$ & $-3.38^{+1.5}_{-0.83}$
& $-1.54^{+1.4}_{-0.43}$ & $-1.46^{+1.3}_{-0.40}$
& $-1.38^{+1.3}_{-0.40}$ & $-1.53^{+1.4}_{-0.43}$ \\

\hline

$\chi^2_{\mathrm{min}}$ 
& $44.38$ & $1595.43$ 
& $38.22$ & $1588.22$ 
& $32.96$ & $1585.56$ \\

$\mathrm{AIC}$ 
& $50.38$ & $1603.4329$
& $52.23$ & $1604.2299$ 
& $48.97$ & $1603.5777$ \\

\hline\hline
\end{tabular} }

\caption{Marginalized $68\%$ constraints on cosmological and model parameters for $\Lambda$CDM, minimal viscous, and non-minimal viscous dark energy models. Each model is shown for both Union3 (U3) and PantheonPlus+SH0ES (P+) supernova datasets combined with DESI DR2 and Planck 2018.}

\label{tab:combined_visc}
\end{table*}

\begin{figure*}
    \centering
    \includegraphics[width=0.45\linewidth]{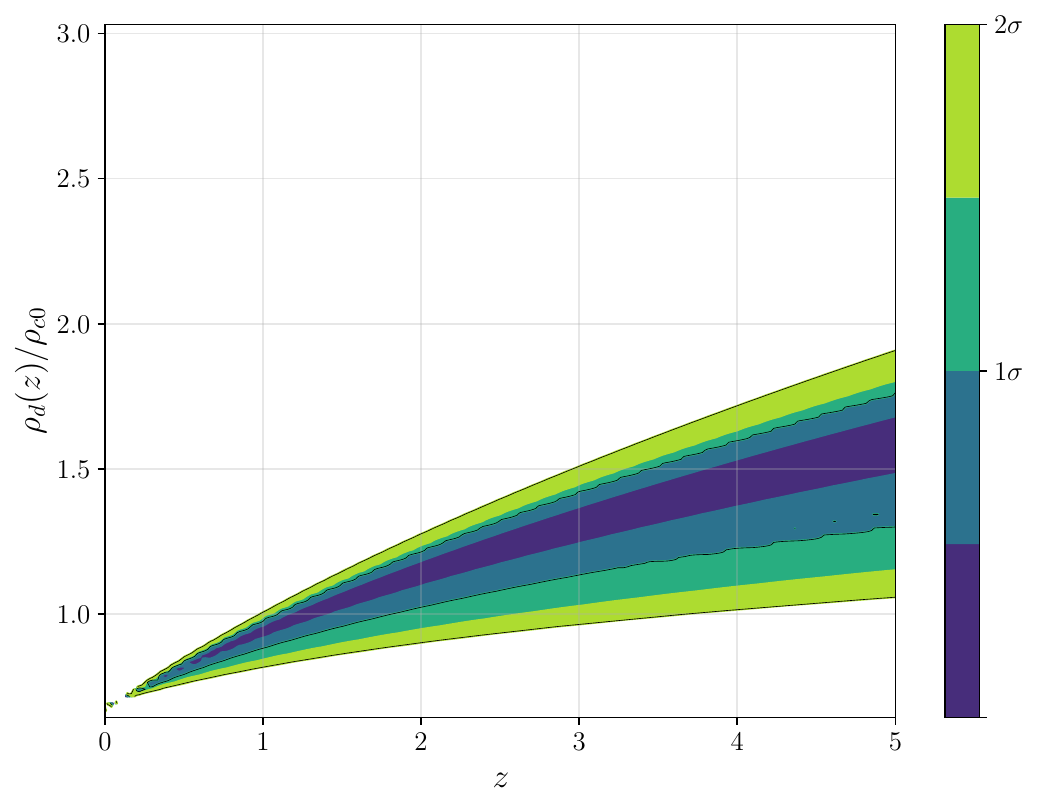}
    \includegraphics[width=0.45\linewidth]{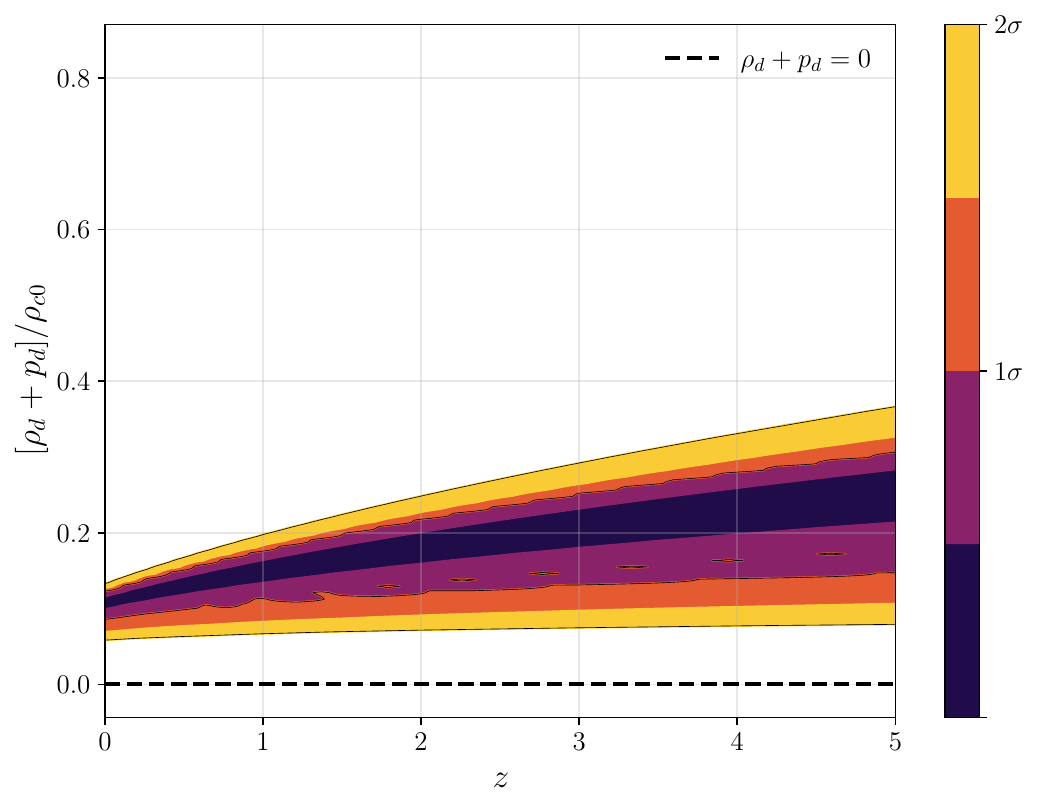} \\
    \includegraphics[width=0.45\linewidth]{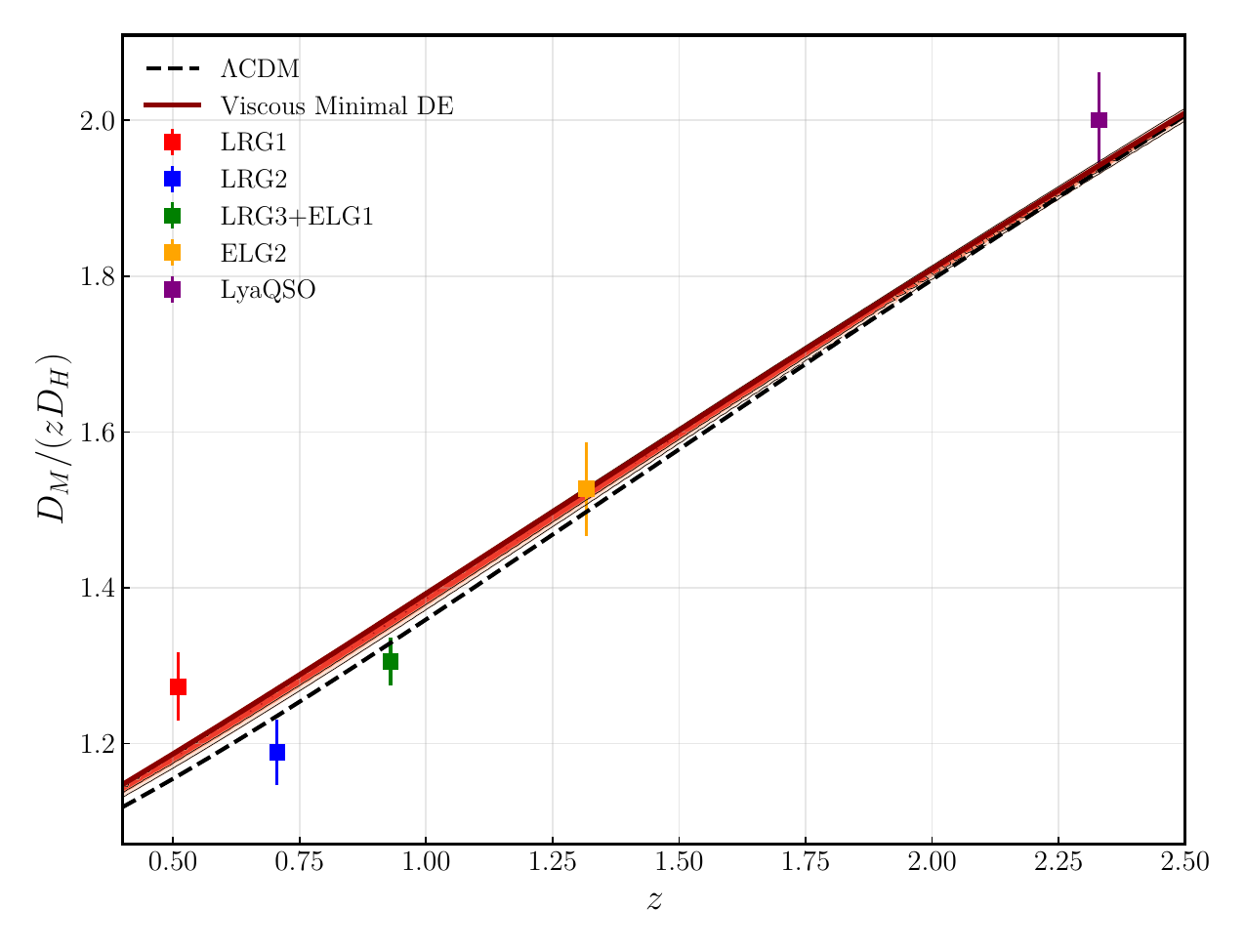}
    \includegraphics[width=0.45\linewidth]{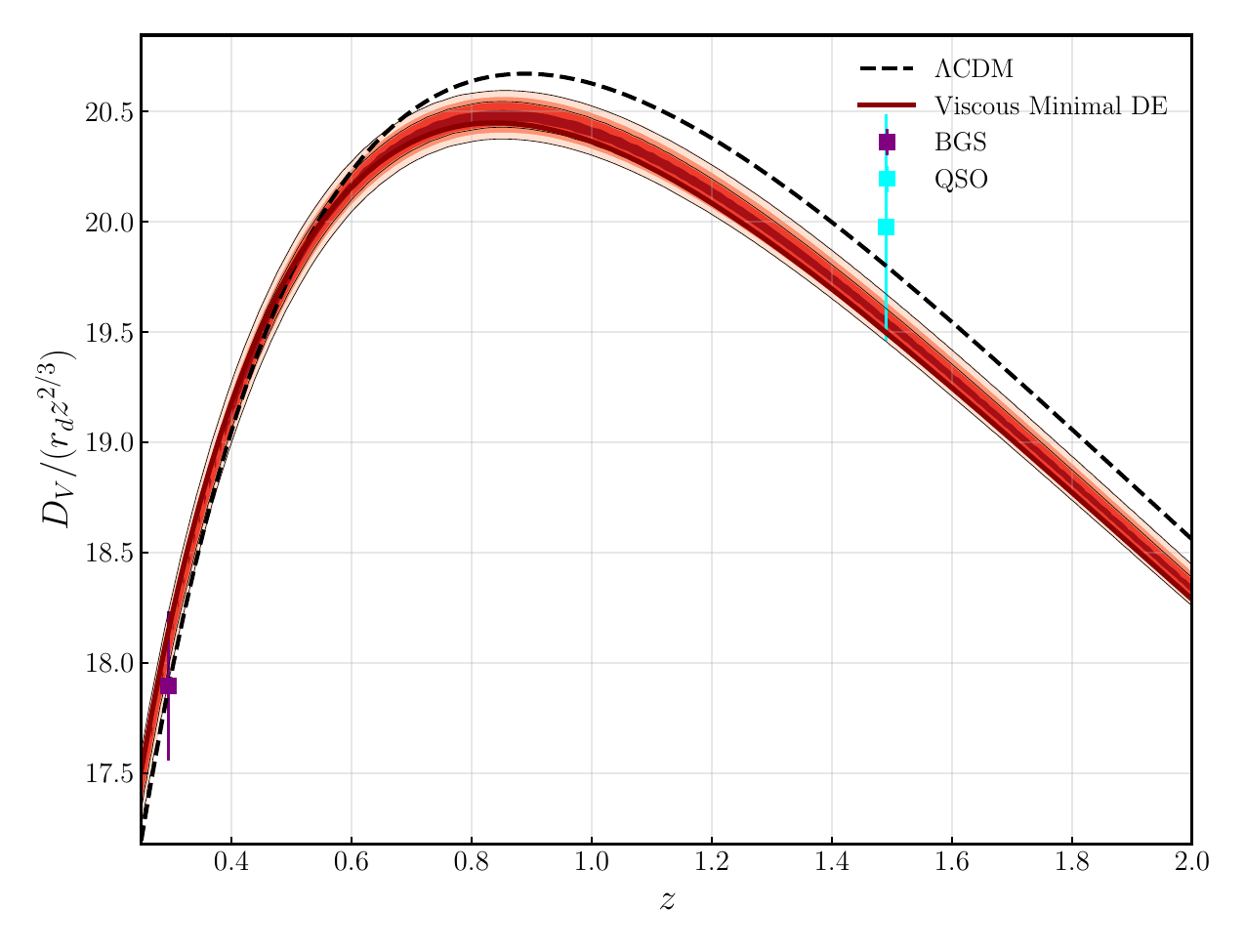}
    \caption{$2\sigma$ Reconstruction of $\rho$, $\rho+p$ and DR2 observables for the minimal case with Union3+DR2+CMB dataset.}
    \label{fig:reconstruct_minimal}
\end{figure*}

In this section, we analyze the minimal coupling case of the viscous dark energy model using combined datasets consisting of either Union3 (U3) or PantheonPlus+SH0ES (P+) supernova compilations together with DESI DR2 BAO measurements and Planck 2018 CMB constraints. The corresponding two-dimensional confidence contours and marginalized posterior distributions are shown in figure~\ref{fig:contour_minimal1}. The posterior shows significant constraints for the cosmological and the model parameters with differentiable correlations. For example, as shown in the figure of the $\Omega_m-\gamma$ panel, we find a negative correlation of $\gamma$ with $\Omega_m$ as is the case of other parameters where we find positive and negative correlations among different set of parameters. The constraints on the cosmological and model parameters are summarized in Table~\ref{tab:combined_visc}. 

In the minimal coupling scenario, we vary the standard cosmological parameters $\Omega_m$, $\Omega_b h^2$, and $h$, together with the viscous model parameters $w$, $\gamma$, $\alpha$, and $n$. For the Union3 combination, we find that the inferred cosmological parameters remain broadly consistent with the expectations from \textit{Planck} 2018~\cite{Planck:2018vyg}, although noticeable shifts are present. In particular, the minimal model favors a larger matter density, $\Omega_m = 0.3257 \pm 0.0062$, together with a lower Hubble parameter, $h = 0.6612^{+0.0056}_{-0.0067}$, relative to $\Lambda$CDM. This behavior reflects the impact of the viscous sector in modifying the late-time expansion history. The dark-energy equation of state is constrained to $w = -0.853^{+0.035}_{-0.021}$, indicating a deviation from the cosmological constant scenario. Other model parameters are also constrained, with $\gamma = -0.0023 \pm 0.0010$ remaining consistent with zero within $1\sigma$, while $\alpha$ is constrained to be very small, $\alpha \sim \mathcal{O}(10^{-5})$, implying a weak viscous contribution. The parameter $n$ remains moderately constrained, allowing some flexibility in the scaling behavior of the viscous component.

When the PantheonPlus+SH0ES compilation is used instead of Union3, the cosmological parameters shift closer to the standard $\Lambda$CDM values. In particular, the Hubble parameter increases to $h = 0.6906 \pm 0.0051$, while the matter density decreases to $\Omega_m = 0.3012 \pm 0.0047$, in better agreement with local distance ladder measurements. Simultaneously, the equation-of-state parameter approaches the cosmological constant limit, with $w = -0.988 \pm 0.023$. Other parameters remain small, with $\gamma = -0.0014 \pm 0.0010$ and $\alpha = (0.61^{+0.21}_{-0.57}) \times 10^{-5}$, indicating that the P+ dataset strongly suppresses large viscous deviations from $\Lambda$CDM. The parameter $n$ shifts toward slightly larger values, $n = 0.543^{+0.25}_{-0.088}$, although it remains only moderately constrained.

From a statistical perspective, the minimal coupling model improves the fit relative to $\Lambda$CDM for both supernova compilations. For the Union3 dataset, the minimum chi-square decreases from $\chi^2_{\mathrm{min}} = 44.38$ for $\Lambda$CDM to $38.22$ for the minimal viscous model. Similarly, for the PantheonPlus+SH0ES dataset, the fit improves from $\chi^2_{\mathrm{min}} = 1595.43$ to $1588.22$. Nevertheless, the Akaike Information Criterion (AIC) shows that part of this improvement is compensated by the larger number of free parameters. Finally, the DESI DR2 data play an important role in constraining the parameter space, particularly favoring lower values of $H_0$ and tighter bounds on deviations from $\Lambda$CDM, consistent with trends reported in recent large-scale structure analyses.

Finally, we present the $2\sigma$ reconstructions of $\rho(z)$, $\rho(z)+p(z)$, and the BAO observables from the DR2 experiment in Fig.~\ref{fig:reconstruct_minimal} using the Union3+DR2+CMB dataset. Instead of reconstructing the equation-of-state parameter $w(z)$ directly, we reconstruct $\rho(z)+p(z)$, since $w(z)$ becomes singular when $\rho(z)\rightarrow 0$. We selected this dataset because it provided the most significant improvement in the $\chi^2$ fit within our analysis. In the left panel of Fig.~\ref{fig:reconstruct_minimal}, the evolution of $\rho(z)$ exhibits clear deviations from a constant value, indicating departures from the standard $\Lambda$CDM scenario. Correspondingly, the right panel shows that $\rho(z)+p(z)$ also deviates from the dashed line associated with the cosmological constant limit, $w=-1$. The bottom panels display the fits to the DESI DR2 BAO observables. In particular, the improved agreement of the $\Lambda$CDM model with the LRG2 and QSO measurements is reflected in its lower $\chi^2$ value compared to the minimal viscous scenario, as summarized in Table~\ref{tab:combined_visc}.

\begin{figure*}
    \centering
    \includegraphics[width=0.85\linewidth]{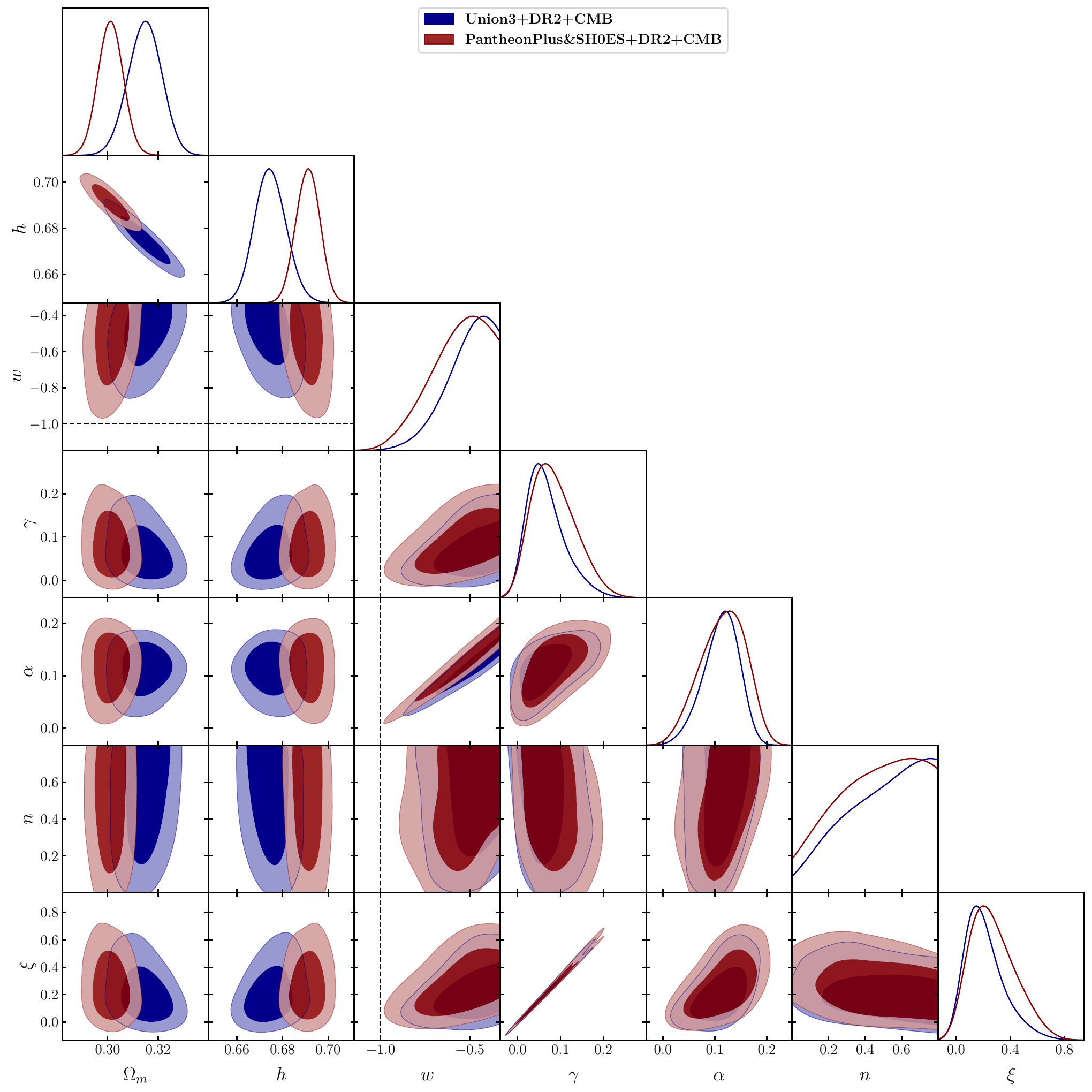}
    \caption{2-D Posteriors with $2\sigma$ confidence levels for the non-minimal case with different SNe Ia along with DR2+CMB dataset. Again, the $\Omega_b h^2$ contours are omitted to reduce the overcrowding of the contours.}
    \label{fig:contour_non-minimal1}
\end{figure*}

\subsection{Case (ii):Non-Minimal Interaction}

We now consider the non-minimal coupling scenario, in which an additional interaction parameter $\xi$ is introduced in the dark  sector. The analysis is performed using the same combined datasets as in the minimal case, namely either Union3 or PantheonPlus+SH0ES supernovae together with DESI DR2 BAO measurements and Planck 2018 CMB constraints. The corresponding two-dimensional confidence contours and marginalized posterior distributions are shown in Fig.~\ref{fig:contour_non-minimal1}. The posterior distributions exhibit significant constraints on both the cosmological and model parameters, with clear parameter degeneracies and correlations, except for the parameter $n$, which remains comparatively weakly constrained. For instance, the $w$–$\alpha$ panel in Fig.\ref{fig:contour_non-minimal1} shows a positive correlation between $\alpha$ and $w$. Similar positive and negative correlations are also observed among other parameter combinations. A more detailed discussion of the parameter degeneracies and correlations is provided in Appendix \ref{appendix:b}. The resulting parameter constraints are summarized in Table~\ref{tab:combined_visc}.

\begin{figure*}[!htbp]
    \centering
    \includegraphics[width=0.45\linewidth]{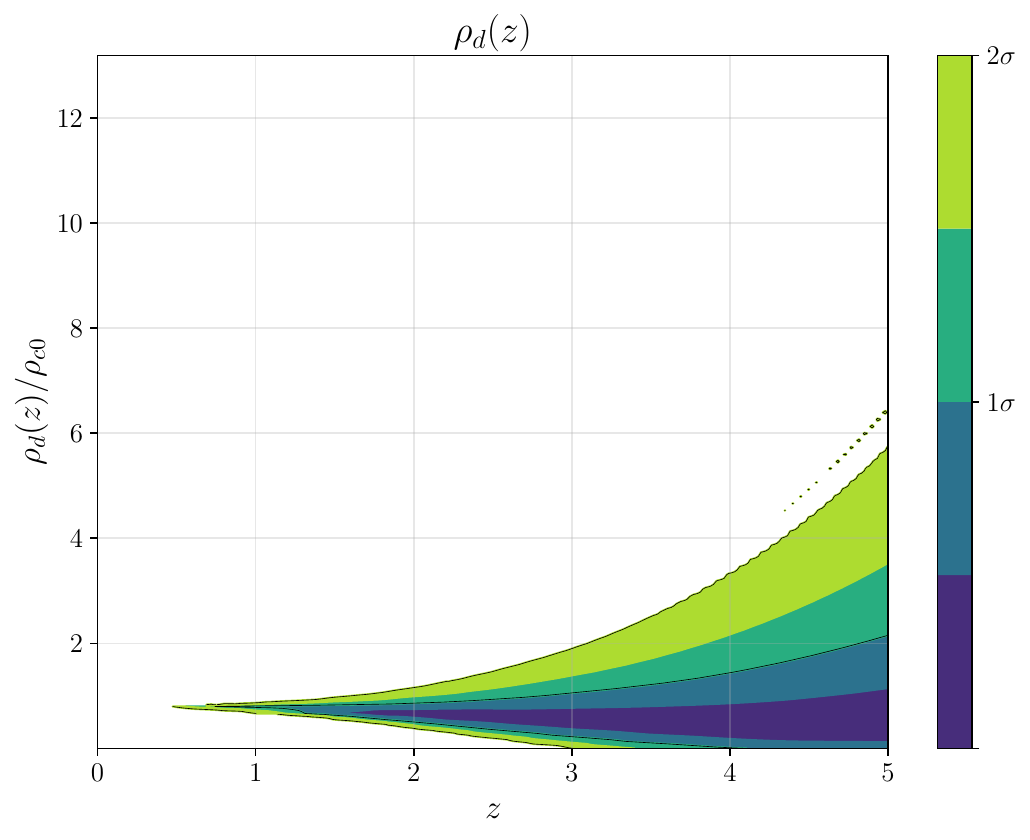}
    \includegraphics[width=0.45\linewidth]{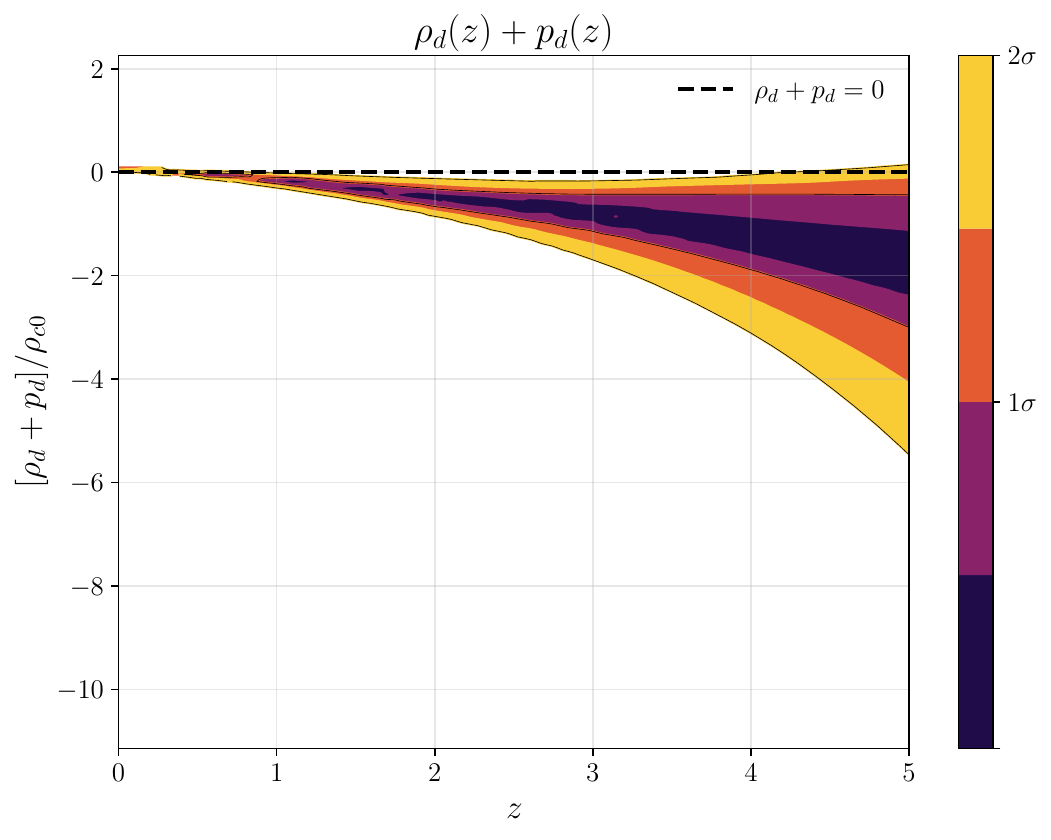} \\

    \includegraphics[width=0.45\linewidth]{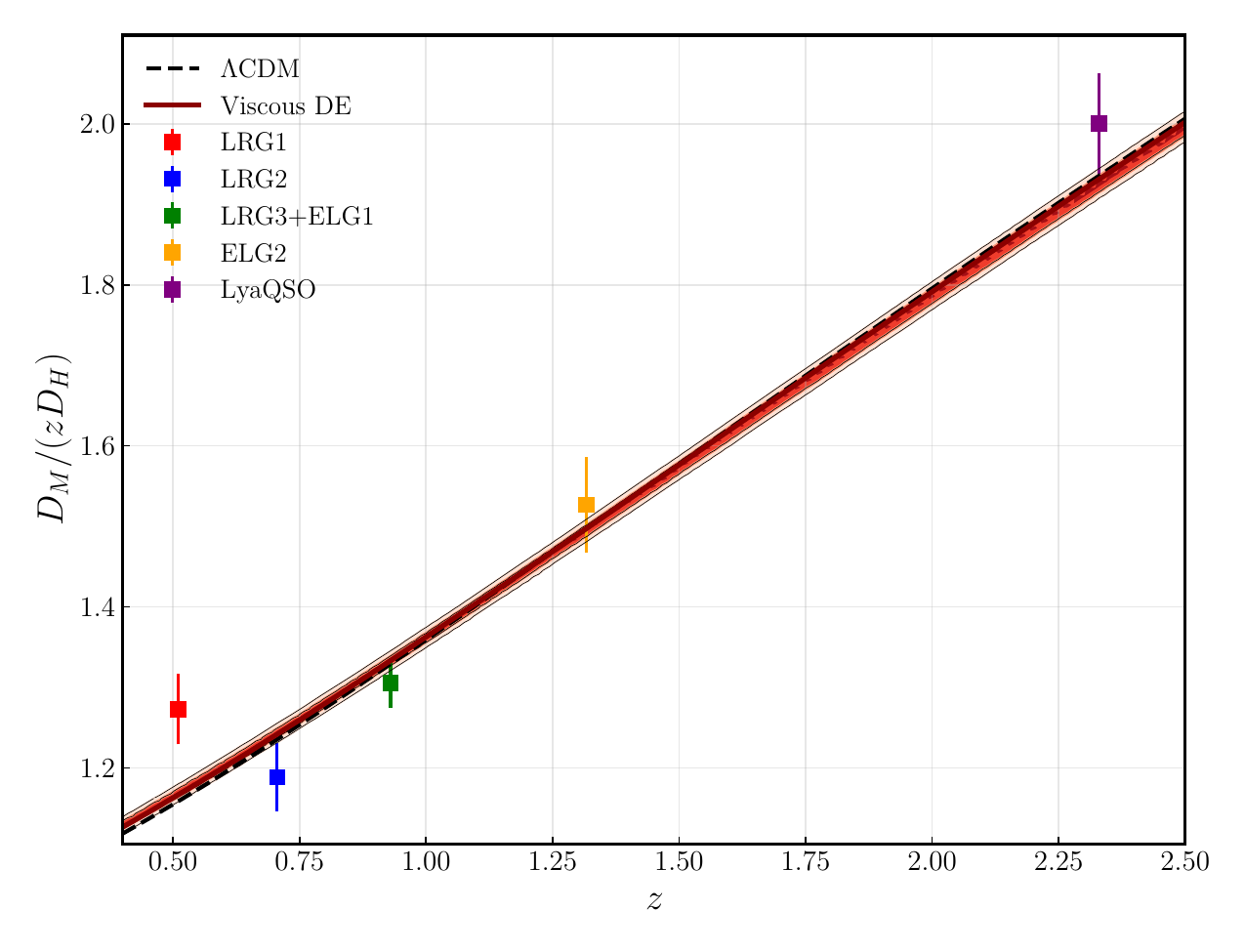}
    \includegraphics[width=0.45\linewidth]{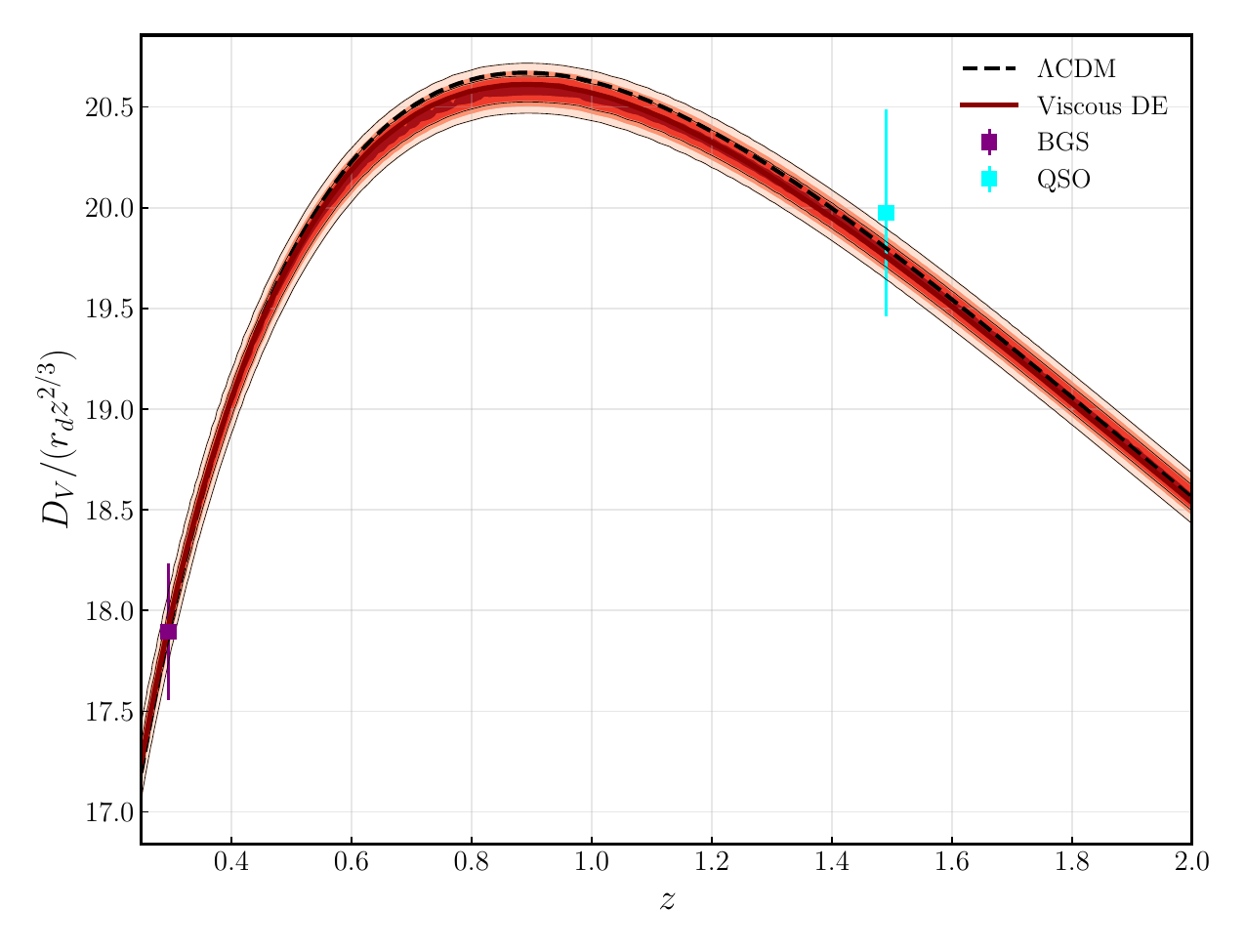}
    \caption{$2\sigma$ Reconstruction of $\rho$, $\rho+p$ and DR2 observables for the non-minimal case with Union3+DR2+CMB dataset. }
    \label{fig:reconstruct_non_minimal}
\end{figure*}

In addition to the standard cosmological parameters $\Omega_m$, $\Omega_b h^2$, and $h$, and the viscous parameters $w$, $\gamma$, $\alpha$, and $n$, the non-minimal model includes the coupling parameter $\xi$. We find that the inclusion of the non-minimal interaction significantly enlarges the allowed parameter space and modifies the cosmological constraints.

For the Union3 combination, the matter density shifts to $\Omega_m = 0.3147 \pm 0.0066$, while the Hubble parameter increases to $h = 0.6747 \pm 0.0068$, moving closer to the $\Lambda$CDM value. The equation of state becomes much less constrained compared to the minimal scenario, with $w = -0.515^{+0.16}_{-0.076}$, indicating a substantial departure from a cosmological constant behavior. Other parameters $\gamma$, $\alpha$, and $n$ also exhibit larger deviations and broader uncertainties. In particular, $\alpha$ increases to $\mathcal{O}(10^{-1})$, several orders of magnitude larger than in the minimal model, while $n$ shifts toward larger values. The coupling parameter itself is constrained to $\xi = 0.207^{+0.077}_{-0.17}$, indicating a preference for a non-zero interaction at approximately the $1\sigma$ level.

For the PantheonPlus+SH0ES combination, the parameter constraints exhibit similar trends. The Hubble parameter increases further to $h = 0.6912 \pm 0.0052$, while the matter density remains close to the $\Lambda$CDM expectation, $\Omega_m = 0.3012 \pm 0.0049$. The equation of state remains far from the cosmological constant limit, with $w = -0.57^{+0.19}_{-0.11}$. Other model parameters continue to favor comparatively large values, with $\gamma = 0.084^{+0.038}_{-0.063}$ and $\alpha = 0.116^{+0.050}_{-0.038}$, while the scaling parameter is constrained to $n = 0.46^{+0.28}_{-0.16}$. The interaction parameter shifts to $\xi = 0.27^{+0.12}_{-0.21}$, again mildly favoring a non-zero coupling.

From a statistical standpoint, the non-minimal model provides the best overall fit among the scenarios considered. For the Union3 dataset, the minimum chi-square improves significantly to $\chi^2_{\mathrm{min}} = 32.96$, while for PantheonPlus+SH0ES it decreases to $\chi^2_{\mathrm{min}} = 1585.56$. Moreover, the Akaike Information Criterion also favors the non-minimal scenario, particularly for the Union3 compilation, indicating that the improvement in fit quality justifies the increased model complexity. Overall, these results suggest that allowing for a non-minimal interaction in the viscous dark energy sector can substantially improve the description of the combined cosmological datasets, while simultaneously producing noticeable shifts in both cosmological and phenomenological parameters.

Finally, we present the $2\sigma$ reconstructions of $\rho(z)$, $\rho(z)+p(z)$, and the BAO observables from the DR2 experiment in Fig.~\ref{fig:reconstruct_non_minimal} using the Union3+DR2+CMB dataset for the non-minimal coupling scenario. Again, We selected this dataset because it provided the most significant improvement in the $\chi^2$ fit within our analysis. In the left panel of Fig.~\ref{fig:reconstruct_non_minimal}, the reconstructed evolution of $\rho(z)$ exhibits significant deviations from a constant behavior, indicating a clear departure from the standard $\Lambda$CDM scenario. Similarly, the right panel shows that $\rho(z)+p(z)$ departs noticeably from the dashed line corresponding to the cosmological constant limit, $w=-1$, reflecting the dynamical nature of the dark-energy sector in the non-minimal model also. The bottom panels display the fits to the DESI DR2 BAO observables. In contrast to the minimal scenario, the non-minimal model provides a significantly improved fit to several BAO measurements, leading to the best overall $\chi^2$ among the models considered. In particular, the improved agreement with the DR2 observables is reflected in the lower value of $\chi^2_{\mathrm{min}}$ compared to both the $\Lambda$CDM and minimal viscous scenarios, as summarized in Table~\ref{tab:combined_visc}.

\section{Conclusions} \label{sec:conclusion}

In this study, we investigated the cosmological implications of bulk viscous fluid dynamics in the context of both minimal and non-minimal coupling scenarios between dark matter and dark energy. The theoretical framework was developed within a spatially flat, homogeneous, and isotropic FLRW universe, where viscous pressure arising from non-equilibrium thermodynamics acts as an effective source capable of driving the late-time accelerated expansion of the Universe. To test the viability of the model, we performed a detailed statistical analysis using recent cosmological observations, including DESI DR2 BAO measurements, Planck 2018 CMB compressed likelihoods, and Type Ia supernova datasets from both Union3 and PantheonPlus+SH0ES.

Our analysis shows that the minimal viscous scenario can successfully reproduce the observed late-time acceleration while remaining broadly consistent with the standard cosmological parameters inferred from $\Lambda$CDM. The inclusion of viscous effects introduces additional freedom in the expansion history without invoking scalar fields or modifications to general relativity. In particular, the viscous contribution modifies the late-time dynamics and allows departures from the cosmological constant behavior. However, although the minimal model improves the overall fit relative to $\Lambda$CDM, the improvement is moderate once the additional model complexity is taken into account.

The non-minimal interacting viscous scenario provides a significantly better description of the combined datasets. In this case, the inclusion of the interaction parameter $\xi$ leads to substantial improvements in the fit quality, yielding the lowest $\chi^2_{\mathrm{min}}$ among all the scenarios considered. Our results indicate that the observational data favor a non-zero interaction parameter, with $\xi$ constrained to positive values at approximately the $2\sigma$ level for both Union3 and PantheonPlus+SH0ES combinations. This suggests that the cosmological observations significantly support an interacting viscous dark-energy scenario, where energy exchange between dark matter and dark energy may play a non-negligible role in the late-time evolution of the Universe.

We further find that the non-minimal model allows larger deviations from the standard $\Lambda$CDM cosmology, particularly in the equation-of-state parameter $w$ and the viscous parameter $\alpha$. The reconstructed evolution of $\rho(z)$ and $\rho(z)+p(z)$ also demonstrates clear departures from the cosmological constant behavior, supporting the possibility of dynamical dark energy driven by dissipative effects. Moreover, the improved agreement with DESI DR2 BAO observables indicates that the interacting viscous framework can better accommodate the late-time expansion history probed by recent large-scale structure observations.

An important outcome of this work is the identification of strong parameter degeneracies, particularly involving the scaling parameter $n$. Although the parameter $n$ is moderately constrained by the current datasets, the posterior analysis reveals that it remains significantly degenerate with other cosmological and viscous parameters. This behavior originates from the nonlinear structure of the viscous pressure contribution, where parameters such as $w$, $\alpha$, and $n$ contribute simultaneously to the effective dynamics. The currently used background datasets mainly constrain combinations of expansion-history parameters, which limits the ability to fully break these degeneracies.

Overall, our findings demonstrate that viscous cosmological models provide a physically motivated and observationally viable alternative to the standard cosmological constant paradigm. In particular, the non-minimal interacting viscous scenario emerges as the preferred framework among the models studied in this work. In future investigations, we plan to further explore the parameter degeneracies by incorporating additional cosmological probes, including growth-rate measurements such as $f\sigma_8$, large-scale structure (LSS) observations, and full-shape galaxy clustering analyses. Furthermore, extending the present framework to perturbation-level cosmology will allow us to investigate the impact of viscous interactions on the growth of cosmic structures and may provide stronger constraints capable of breaking the existing degeneracies among the model parameters. Future high-precision datasets from upcoming surveys such as \textit{Euclid} \cite{Amendola:2016saw}, DESI upcoming surveys \cite{DESI:2022lza}, and the Vera C. Rubin Observatory \cite{Shah:2023sna} are expected to provide further insight into the dissipative nature of the dark sector.

\section*{Acknowledgments}
SA acknowledges the CSIR fellowship provided by the Govt. of India under the CSIR-SRF scheme (file no. 09/0466(12904)/2021). SAA acknowledges the support of the DGAPA postdoctoral fellowship program at ICF-UNAM, Mexico. Both SA and SAA also acknowledges the High Performance Computing facility Pegasus at IUCAA, Pune, India. JAV acknowledges support from FOSEC SEP-CONACYT Ciencia B\'asica A1-S-21925,  UNAM-DGAPA-PAPIIT IN109126, IN110325 and Cátedra de Investigación Marcos Moshinsky.

\bibliography{viscous}
\appendix

\section{ Consistency between the analytic and numerical formulations}
\label{appendix:a}

We begin from the evolution equation for the normalized Hubble function $E(a) = H(a)/H_0$ expressed in terms of the scale factor $a$:
\begin{align}
\label{eq:h_a}
a\frac{dE}{da}
&+ \frac{3}{2}(1+w)E
+ \frac{3}{2}(\gamma-\omega)
\frac{\Omega_{dm0}a^{-3(1+\gamma)+\xi}}{E}
- \frac{3}{2}\omega
\frac{\Omega_{b0}a^{-3}}{E}
\nonumber\\
&- \frac{3}{2}\left(\omega-\frac13\right)
\frac{\Omega_{r0}a^{-4}}{E}-\frac{3\sqrt3}{2}\alpha E^{2n}
\nonumber\\
&\Bigg[
1-
\frac{
\Omega_{dm0}a^{-3(1+\gamma)+\xi}
+\Omega_{b0}a^{-3}
+\Omega_{r0}a^{-4}
}{E^2}
\Bigg]^n
=0 .
\end{align}

For numerical purposes it is convenient to rewrite the equation in terms of the logarithmic redshift variable
\begin{equation}
x \equiv \ln(1+z) = -\ln a,
\qquad \Rightarrow \qquad
a = e^{-x}.
\end{equation}

Differentiating, we obtain
\begin{equation}
\frac{da}{dx} = -a
\quad \Rightarrow \quad
a\frac{dE}{da} = -\frac{dE}{dx}.
\label{eq:adx}
\end{equation}

Substituting Eq.~\eqref{eq:adx} into Eq.~\eqref{eq:h_a}, the evolution equation becomes
\begin{align}
\label{eq:h_x}
\frac{dE}{dx}
&= \frac{3}{2}(1+w)E
+ \frac{3}{2}(\gamma-\omega)
\frac{\Omega_{dm0}a^{-3(1+\gamma)+\xi}}{E}
- \frac{3}{2}\omega
\frac{\Omega_{b0}a^{-3}}{E}
\nonumber\\
&\quad
- \frac{3}{2}\left(\omega-\frac13\right)
\frac{\Omega_{r0}a^{-4}}{E}
\nonumber\\
&\quad
-\frac{3\sqrt3}{2}\alpha E^{2n}
\Bigg[
1-
\frac{
\Omega_{dm0}a^{-3(1+\gamma)+\xi}
+\Omega_{b0}a^{-3}
+\Omega_{r0}a^{-4}
}{E^2}
\Bigg]^n .
\end{align}
where $a=e^{-x}$.

Equation~\eqref{eq:h_x} is a first-order, nonlinear ordinary differential equation (ODE) for $E(x)$, where $x = \ln(1+z)$. We solve this ODE numerically using an explicit fourth-order Runge--Kutta integrator (RK4) on a fixed grid in $x$, starting from $x=0$ and stepping toward higher redshift. Each physical contribution in Eq.~\eqref{eq:h_x} appears explicitly in the right-hand side:
\begin{align}
\text{(i)}\ & \frac{3}{2}(1+w)\,E, \\
\text{(ii)}\ & \frac{3}{2}(\gamma-\omega)\,
\frac{\Omega_{dm0}\,a^{-3(1+\gamma)+\xi}}{E}, \\
\text{(iii)}\ & -\frac{3}{2}\omega\,
\frac{\Omega_{b0}\,a^{-3}}{E}, \\
\text{(iv)}\ & -\frac{3}{2}\left(\omega-\tfrac{1}{3}\right)\,
\frac{\Omega_{r0}\,a^{-4}}{E}, \\
\text{(v)}\ & -\frac{3\sqrt{3}}{2}\,\alpha\,E^{2n}
\left[1-\frac{\Omega_{dm0}\,a^{-3(1+\gamma)+\xi}
+\Omega_{b0}\,a^{-3}
+\Omega_{r0}\,a^{-4}}
{E^2}
\right]^n .
\end{align}
These terms correspond respectively to:
(i) the homogeneous part controlled by the effective equation-of-state parameter $w$,
(ii) the interacting/viscous dark matter sector,
(iii) baryons,
(iv) radiation,
and (v) the nonlinear viscous contribution proportional to $\alpha$.

The initial condition is fixed by the normalization at the present epoch,
\begin{equation}
E(x=0) = \frac{H(z=0)}{H_0} = 1,
\end{equation}
so that $E$ coincides with the usual dimensionless expansion rate $E(z)=H(z)/H_0$ today.

After numerically integrating Eq.~\eqref{eq:h_x} to obtain $E(x)$ over the redshift range of interest, we evaluate the expansion history through
\begin{equation}
E(z) \equiv E(x), 
\qquad x = \ln(1+z),
\qquad a = \frac{1}{1+z},
\end{equation}
This is the background expansion rate used by all cosmological likelihoods (e.g. BAO, supernovae, CMB distances) in our analysis.

Therefore, Eq.~\eqref{eq:h_a} and its transformed form Eq.~\eqref{eq:h_x} are fully consistent with the numerical background evolution used in the viscous cosmology fits presented in this work.

\section{Degeneracies among Parameters in the Viscous Scenario}
\label{appendix:b}

\begin{figure*}[!htbp]
    \centering
    \includegraphics[width=0.85\linewidth]{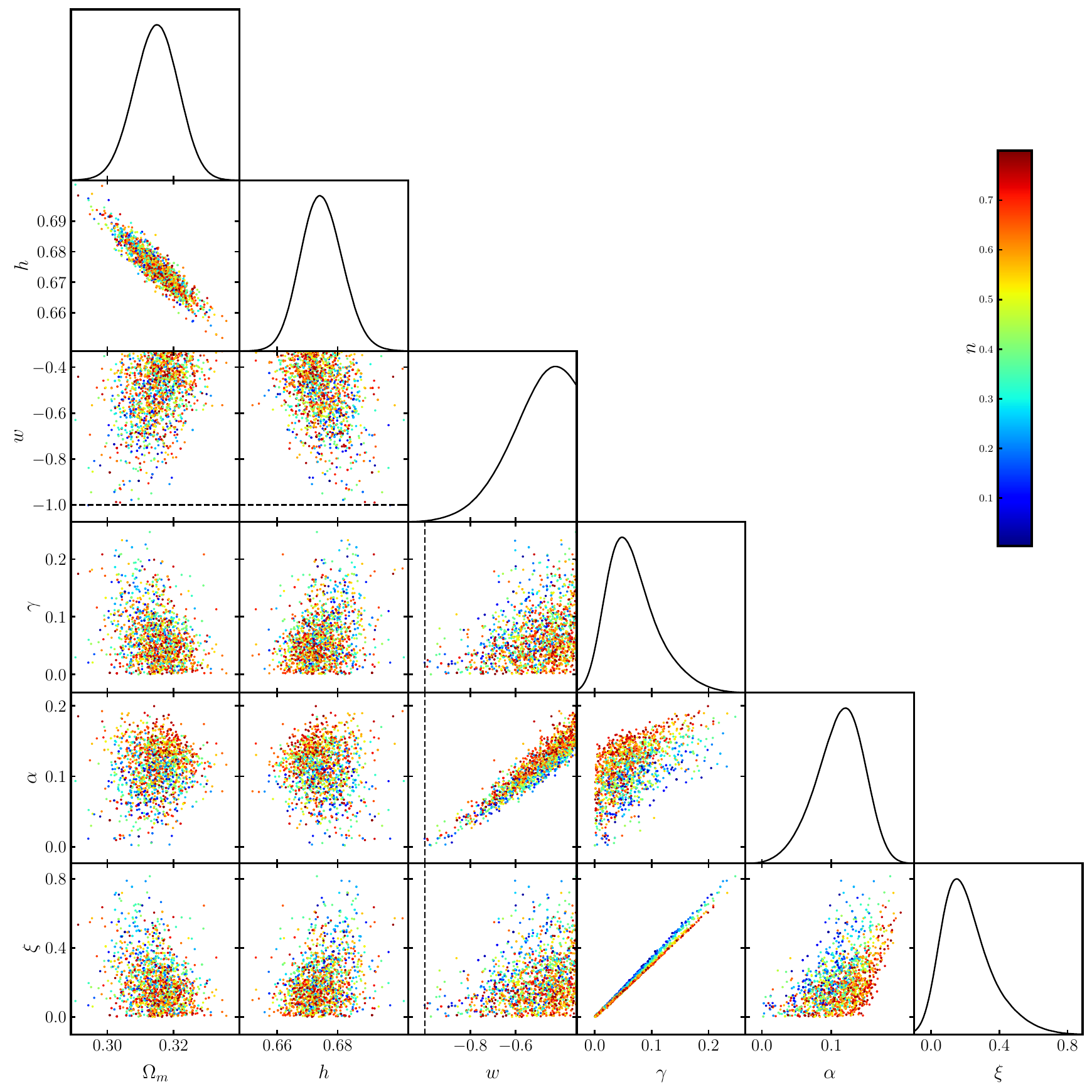}
    \caption{2-D Posteriors with heatmap for the degeneracies with n}
    \label{fig:contour_non-minimal2}
\end{figure*}

\begin{figure*}[!htbp]
    \centering
    \includegraphics[width=0.85\linewidth]{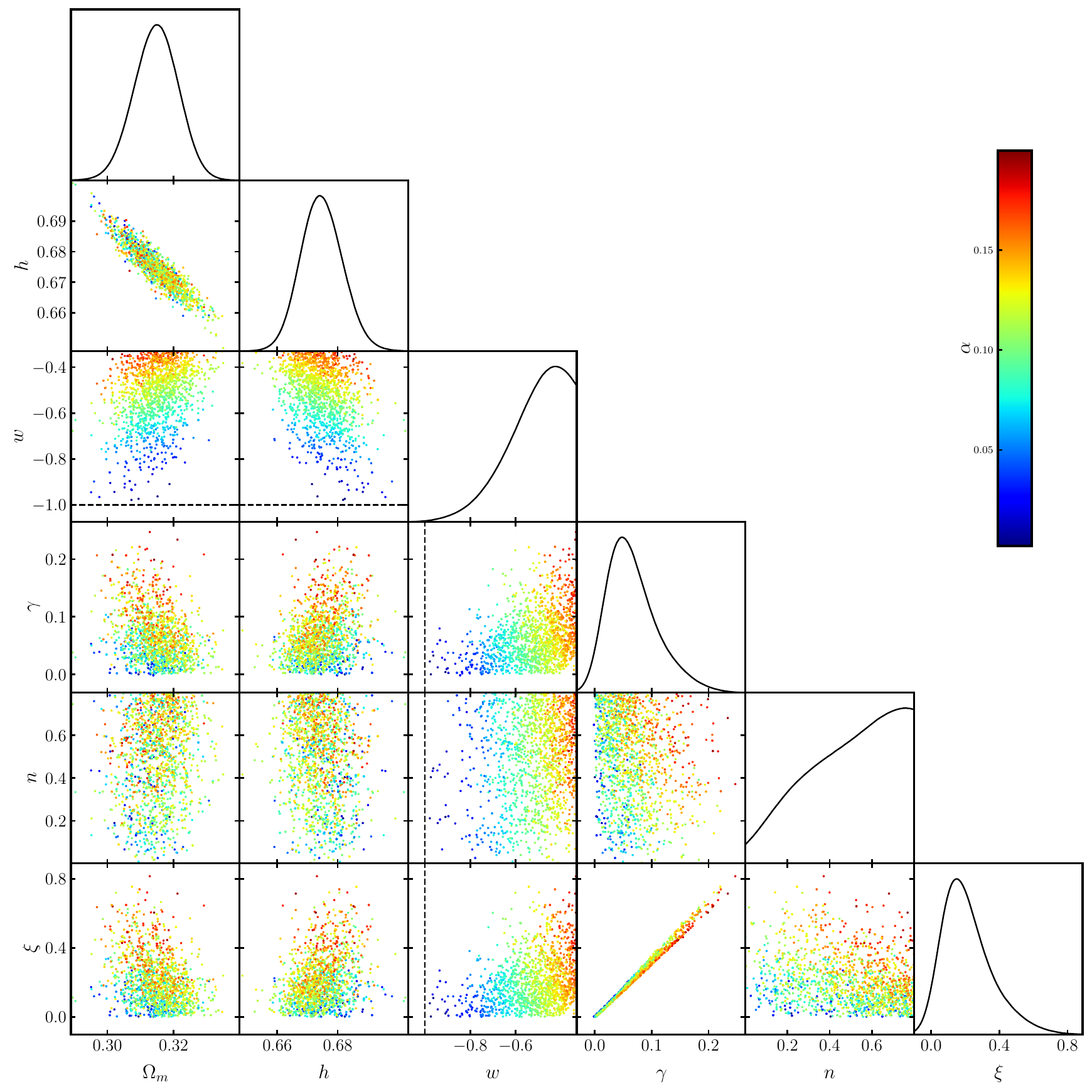}
    \caption{2-D Posteriors with heatmap for the degeneracies with $\alpha$}
    \label{fig:contour_non-minimal3}
\end{figure*}

\begin{figure*}[!htbp]
    \centering
    \includegraphics[width=0.85\linewidth]{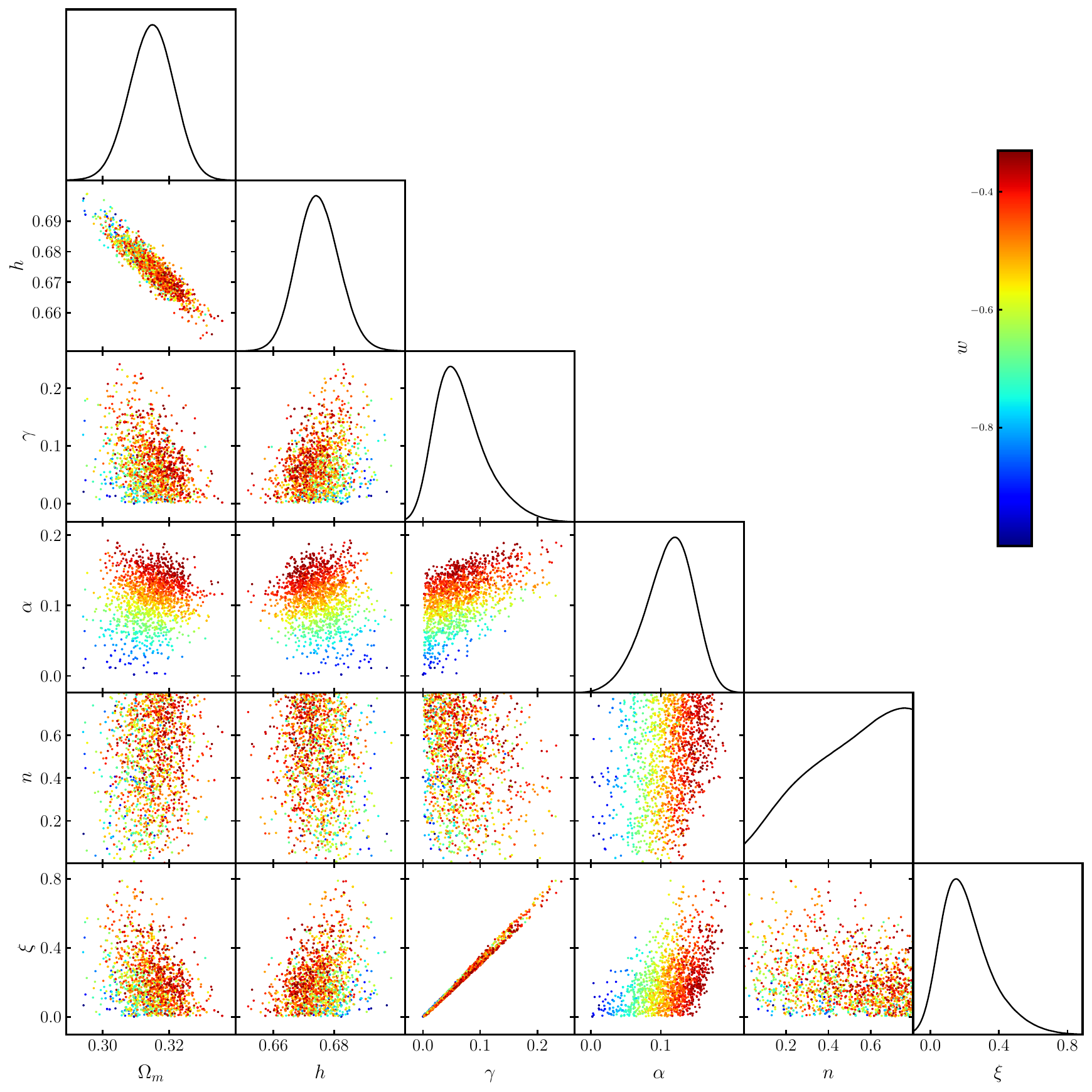}
    \caption{2-D Posteriors with heatmap for the degeneracies with $w$}
    \label{fig:contour_non-minimal4}
\end{figure*}

The introduction of additional parameters in the viscous dark energy model naturally leads to parameter degeneracies. In this section, we discuss the degeneracies and correlations among the parameters of the viscous scenario. In \cref{fig:contour_non-minimal2,fig:contour_non-minimal3,fig:contour_non-minimal4}, we illustrate the degeneracies associated with the parameters $n$, $\alpha$, and $w$, respectively. These degeneracies originate from the relation defined in Eq.~\eqref{eq:pressure_de}, where the effective pressure depends simultaneously on $w$, $\alpha$, and $n$. Consequently, variations in one of these parameters can be compensated by changes in the others, leading to correlated parameter spaces.

In Fig.~\ref{fig:contour_non-minimal2}, we present the effect of the parameter $n$ on different cosmological and model parameters. The heatmap corresponds to the values of $n$, ranging from $0$ (dark blue) to $0.8$ (red). We observe that a broad range of $n$ values produces nearly identical constraints on parameters such as $\Omega_m$, $h$, and other cosmological quantities. This demonstrates the strong degeneracy of $n$ with the remaining parameters. As a consequence, the parameter $n$ is only moderately constrained by the datasets considered in this work, since the observations primarily constrain combinations of parameters related to $\Omega_m$, $h$, and $w$.

In Fig.~\ref{fig:contour_non-minimal3}, we show the heatmap associated with the parameter $\alpha$, varied within the range $[0.00,0.20]$. The figure reveals that the posterior distributions are dominated by low and intermediate values of $\alpha$, whereas the high-$\alpha$ regions (red colors) correspond to significantly lower posterior density. This indicates a clear correlation between $\alpha$ and the other cosmological parameters. For example, in the $\Omega_m$--$w$ plane, with $\alpha$ represented as the third dimension through the heatmap, distinct color patterns appear, illustrating the correlated behavior among these parameters.

Finally, in Fig.~\ref{fig:contour_non-minimal4}, we illustrate the degeneracy associated with the equation-of-state parameter $w$. The heatmap representation shows that regions corresponding to less negative values of $w$ dominate the posterior distributions, while the blue regions associated with values closer to the standard $\Lambda$CDM limit, $w=-1$, are comparatively less favored. This behavior further supports the results discussed in the main analysis section. As an example, in the $\xi$--$\gamma$ plane, using $w$ as the third dimension in the heatmap, the dominance of the red region indicates a preference for lower negative values of $w$ within the non-minimal viscous scenario. Similar behavior is also seen in $\alpha-\gamma$ and $\alpha-\Omega_m$ planes which is consistent with the arguments for $\xi$--$\gamma$ plane.

\end{document}